\documentclass[pdflatex, sn-mathphys-num]{sn-jnl}

\usepackage{graphicx} 
\usepackage{pgfplots}
\pgfplotsset{compat=1.18}
\usepgfmodule{nonlineartransformations}
\usepackage{tikz}
\usetikzlibrary{decorations, arrows.meta, calligraphy, curvilinear,shapes.geometric,shapes.arrows}

\usepackage{multirow}%
\usepackage{amsmath,amssymb,amsfonts}%
\usepackage{amsthm}%
\usepackage{mathrsfs}%
\usepackage[title]{appendix}%
\usepackage{xcolor}%
\usepackage{textcomp}%
\usepackage{manyfoot}%
\usepackage{booktabs}%
\usepackage{algorithm}%
\usepackage{algorithmicx}%
\usepackage{algpseudocode}%
\usepackage{listings}%
\usepackage{subcaption}

\usepackage{bm} 

\usepackage{comment}
\usepackage{subcaption}
\usepackage{mathtools}
\usepackage{siunitx}   

\usepackage{xcolor}
\definecolor{TUDa-0D}{cmyk/RGB/HTML}{0,0,0,.8/83,83,83/535353}
\definecolor{TUDa-0c}{cmyk/RGB/HTML}{0,0,0,.6/137,137,137/898989}
\definecolor{TUDa-0b}{cmyk/RGB/HTML}{0,0,0,.4/181,181,181/B5B5B5}
\definecolor{TUDa-0a}{cmyk/RGB/HTML}{0,0,0,.2/220,220,220/DCDCDC}
\definecolor{TUDa-10c}{cmyk/RGB/HTML}{.5,1,.3,0/149,17,105/951169}

\makeatletter
\def\mytransformation{%
\pgfmathsetmacro{\myX}{\pgf@x+0.008*\pgf@y*\pgf@y-0.048*\pgf@y}
\pgfmathsetmacro{\myY}{0.0005*\pgf@x*\pgf@x+\pgf@y}
\setlength{\pgf@x}{\myX pt}
\setlength{\pgf@y}{\myY pt}
}
\makeatother

\usepackage{glossaries}
\newacronym{ad}{AD}{automatic differentiation}
\newacronym{adam}{ADAM}{adaptive moment estimation}
\newacronym{bvp}{BVP}{boundary value problem}
\newacronym{cnn}{CNN}{convolutional neural network}
\newacronym{cad}{CAD}{computer-aided design}

\newacronym{dft}{DFT}{discrete Fourier transform}
\newacronym{dof}{d.o.f.}{degrees of freedom}
\newacronym{dl}{DL}{deep learning}

\newacronym{rnn}{RNN}{recurrent neural network}
\newacronym{fe}{FE}{finite element}
\newacronym{fem}{FEM}{finite element method}
\newacronym{feti}{FETI}{finite element tearing and interconnecting}

\newacronym{bc}{BC}{boundary condition}
\newacronym{bem}{BEM}{boundary element method}
\newacronym{fv}{FV}{finite volume}
\newacronym{fc}{FC}{fully connected}
\newacronym{iga}{IGA}{isogeometric analysis}
\newacronym{lbfgs}{L-BFGS}{limited-memory Broyden-Fletcher-Goldfarb-Shanno}
\newacronym{lstm}{LSTM}{long-short term memory}
\newacronym{ml}{ML}{machine learning}
\newacronym{mc}{MC}{Monte Carlo}
\newacronym{nn}{NN}{neural network}
\newacronym{nurbs}{NURBS}{non-uniform rational B-spline}
\newacronym{ode}{ODE}{ordinary differential equation}
\newacronym{piml}{PIML}{physics-informed machine learning}
\newacronym{pinn}{PINN}{physics-informed neural network}
\newacronym{pde}{PDE}{partial differential equation}
\newacronym{relu}{ReLU}{rectified linear units}
\newacronym{resnet}{ResNet}{residual neural network}
\newacronym{sgd}{SGD}{stochastic gradient descent}
\newacronym{rmse}{RMSE}{root mean squared error}
\newacronym{tl}{TL}{transfer learning}

\DeclareMathOperator*{\argmin}{argmin}

\newcommand{\vecpot}{\mathbf{a}}

\newcommand{\GN}{\Gamma_{\textrm{N}}}
\newcommand{\GD}{\Gamma_{\textrm{D}}}

\begin{document}

\title[Multi-patch isogeometric neural solver for partial differential equations on computer-aided design domains]{Multi-patch isogeometric neural solver for partial differential equations on computer-aided design domains}


\author[1]{\fnm{Moritz} \sur{von Tresckow}}\email{moritz.von\_tresckow@tu-darmstadt.de}
\equalcont{These authors contributed equally to this work.}

\author[2]{\fnm{Ion Gabriel} \sur{Ion}}\email{igi@terraquantum.swiss}
\equalcont{These authors contributed equally to this work.}

\author*[3]{\fnm{Dimitrios} \sur{Loukrezis}}\email{d.loukrezis@cwi.nl}

\affil[1]{\orgdiv{Institute for Accelerator Science and Electromagnetic Fields}, \orgname{Technische Universit\"at Darmstadt}, \orgaddress{\street{Schlossgartenstr. 8}, \city{Darmstadt}, \postcode{64289}, \country{Germany}}}

\affil[2]{\orgname{Terra Quantum AG}, \orgaddress{\street{Kornhausstr. 25}, \city{St. Gallen}, \postcode{9000}, \country{Switzerland}}}

\affil*[3]{\orgdiv{Scientific Computing}, \orgname{Centrum Wiskunde \& Informatica}, \orgaddress{\street{Science Park 123}, \city{Amsterdam}, \postcode{1098 XG}, \country{The Netherlands}}}

\abstract{
This work develops a computational framework that combines physics-informed neural networks with multi-patch isogeometric analysis to solve partial differential equations on complex computer-aided design geometries. 
The method utilizes patch-local neural networks that operate on the reference domain of isogeometric analysis. 
A custom output layer enables the strong imposition of Dirichlet boundary conditions.
Solution conformity across interfaces between non-uniform rational B-spline patches is enforced using dedicated interface neural networks. 
Training is performed using the variational framework by minimizing the energy functional derived after the weak form of the partial differential equation. 
The effectiveness of the suggested method is demonstrated on two highly non-trivial and practically relevant use-cases, namely, a 2D magnetostatics model of a quadrupole magnet and a 3D nonlinear solid and contact mechanics model of a mechanical holder. 
The results show excellent agreement to reference solutions obtained with high-fidelity finite element solvers, thus highlighting the potential of the suggested neural solver to tackle complex engineering problems given the corresponding computer-aided design models.
}




\maketitle

\section{Introduction}
\label{sec:intro}


\subsection{Motivation}
In recent years, \glspl{pinn} \cite{raissi2019physics} have emerged as a possible mesh-free alternative to traditional, mesh-based numerical methods for the solution of \glspl{pde}.
Within the \gls{pinn} framework, \glspl{nn} are trained to learn a \gls{pde} solution by minimizing a loss function which incorporates the \gls{bvp} that describes the underlying physics of the system under investigation \cite{anitescu2023physics, kim2024review}.  
It must be noted that methods based on very similar ideas were developed prior to the seminal \gls{pinn} paper, such as the deep Ritz and deep Galerkin methods \cite{e2018deep, sirignano2018dgm}.
However, \gls{pinn} has been established as a blanket term for most related approaches and is used as such within this paper as well.

The \gls{pinn} framework and its numerous variants \cite{yang2021b, kharazmi2021hp, chiu2022can, zhu2023bc, su2024pinn} have already found applications in several fields of computational science and engineering, as in computational electromagnetics \cite{khan2022physics, beltran2022physics, baldan2023physics, lippert2024transfer}, fluid dynamics \cite{cai2021physics_heat, cai2021physics_fluid, eivazi2022physics, zhao2024comprehensive}, and mechanics \cite{samaniego2020energy, bai2023physics, linden2023neural, hu2024physics}, to name only a few indicative application areas.
Nonetheless, due to a number of weaknesses, \glspl{pinn} cannot yet be regarded as competitive to traditional numerical solvers \cite{markidis2021old, grossmann2024can}.
This work focuses on the severe difficulties that \glspl{pinn} face when applied to complex domains, in particular shapes and geometries stemming from \gls{cad}. In turn, these difficulties hinder the application of \glspl{pinn} to real-world engineering problems, where complex, multi-patch \gls{cad} geometries are ubiquitous.

In this context, three main challenges remain largely unaddressed. 
The first challenge concerns the enforcement of solution conformity across complex, multi-patch domains. 
Solution continuity in particular is crucial when dealing with different materials across domain interfaces.
The second challenge concerns the imposition of Dirichlet \glspl{bc}. 
The standard \gls{pinn} framework and most related approaches use a soft-constraints approach, such that the \gls{pinn} learns the \glspl{bc} during training. However, in many settings, the \glspl{bc} must be strongly conformed to. 
The third challenge concerns handling geometry parts with significant variations in size and shape.
Such differences make data normalization difficult, especially under physical conformity requirements.

\subsection{Contribution}
This work aims to enable the application of \glspl{pinn} to \gls{cad} geometries by addressing the three aforementioned limitations.
Of particular interest are complex, multi-patch \gls{cad} geometries, which are rarely addressed in the relevant literature.  
To that end, we introduce a novel neural solver which combines \glspl{pinn} with \gls{iga} \cite{hughes2005isogeometric, cottrell2009isogeometric, nguyen2015isogeometric}. 
More specifically, the neural solver developed in this work is rooted in the \gls{feti} method \cite{farhat1991method} and its extension to \gls{iga} \cite{kleiss2012ieti}, as well as on mortaring methods \cite{apostolatos2014nitsche, brivadis2015isogeometric, merkel2022torque}. 
Therein, the \gls{pinn} framework is integrated.

Our approach consists of developing \gls{nn}-based ansatz functions that approximate the \gls{pde} solution by operating on the reference domain of \gls{iga}.
The solution on the physical domain is obtained by using the pushforward map from the reference to the physical domain, as given by the \gls{nurbs}-based parametrization of the geometry, similar to the conventional \gls{iga}.
The suggested ansatz functions also satisfy Dirichlet \glspl{bc} where necessary, ensure differentiability within the subdomains (patches) of the physical geometry, and ensure solution continuity across patches.

One set of ansatz functions is used to approximate the \gls{pde} solution within the reference domain of each individual patch, thereby leveraging the geometric accuracy of \gls{iga}. 
The output layer of these \glspl{nn} is modified to strongly impose the Dirichlet boundary conditions where necessary. 
Another set of ansatz functions is defined on the interfaces between neighboring patches, which appropriately connect the solutions of the individual patches, thus ensuring solution conformity across them.
Training is performed by utilizing the variational \gls{pinn} framework \cite{e2018deep, kharazmi2021hp, khodayi2020varnet}, where the loss function to be minimized is the discretized energy functional derived after the weak form of the \gls{pde}. 
This is a natural choice, as it aligns with the principles of \gls{iga}, which is also built upon variational formulations.

The suggested neural solver offers three distinct contributions, each connected to one of the challenges identified above. 
First, it ensures control over solution continuity across patch interfaces. 
Second,
it enables the strong imposition of Dirichlet \glspl{bc}. 
Third, by operating on the reference domain, it naturally normalizes \gls{nn} inputs.
The efficacy of the method is demonstrated on two real-world engineering applications. 
The first use-case is drawn from the field of computational electromagnetics and concerns a 2D magnetostatics model of a quadrupole magnet.
The second use-case is drawn from the field of computational mechanics and concerns a 3D nonlinear solid mechanics model of a mechanical holder.
For the latter, we employ a hyperelastic material model combined with contact \glspl{bc} and introduce parametric dependencies in the solution.

\subsection{Related work}
Prior works have explored the combination of \glspl{nn} and \glspl{pinn} with \gls{iga}, e.g., for applications in solid mechanics \cite{goswami2020transfer, haghighat2021physics, cao2025parametric} or electromagnetics \cite{von2022neural, kostas2024physics}, including \gls{iga} variants such as isogeometric collocation methods \cite{moller2021physics} and the isogeometric boundary element method \cite{kostas2024physics}. 
Differentiable \gls{nurbs} \cite{prasad2022nurbs} were developed as a means to integrate \gls{cad} models into deep learning frameworks.
Isogeometric \glspl{nn} \cite{gasick2023isogeometric} combine \gls{nurbs} with \glspl{nn} to predict \gls{nurbs} coefficients and approximate \gls{pde} solutions for varying physical and geometric parameters. 
Deep \glspl{nn} have been used to determine the optimum number of quadrature points for evaluating \gls{iga} stifness matrices \cite{nath2025transfer}.
The deep \gls{nurbs} method \cite{saidaoui2024deep} introduces an ansatz for boundary-conforming \glspl{nn}. 
Graph \glspl{nn} have been employed to learn the \gls{pde} solution at the control points of splines \cite{li2023isogeometric, xu2025iga}, while convolutional \glspl{nn} have also been used in a similar context \cite{wang2022iga, lu2023convolution, zhang2023isogeometric, zhang2025multi}.
However, most, if not all, of these related works concern single-patch geometries. 
Hence, their application for complex, multi-patch domains is not straightforward. 
For the same reason, solution conformity across interfaces is rarely a concern. 
One exception is the work of von Tresckow et al. \cite{von2022neural}, which does consider multi-patch \gls{cad} geometries and introduces a discontinuous Galerkin-based \gls{pinn} formulation to weakly impose interface conditions and Dirichlet \glspl{bc}. 
The approach developed in this work is substantially different. 

In relation to solution conformity across interfaces, prior works have combined \glspl{pinn} with domain decomposition methods, using different strategies such as physics-based coupling terms \cite{jagtap2020extended, jagtap2020conservative}, network combinations \cite{zhang2022multi}, or tailored \gls{nn} architectures \cite{sarma2024interface}. 
In contrast to the present work, these methods are not applied in the context of \gls{cad} and consider relatively simple computational domains. 
Additionally, the present work assigns interface-specific \glspl{nn} to enforce the correct solution continuity, which is a new approach, at least to the knowledge of the authors.

Concerning the imposition of Dirichlet \glspl{bc}, most approaches based on the \gls{pinn} framework enforce \glspl{bc} only weakly, by incorporating them into the loss function.
A few works have considered the exact imposition of Dirichlet \glspl{bc}, mainly by means of dedicated distance functions
\cite{sukumar2022exact, wang2023exact}. 
In contrast, our method employs a custom output layer with residual connection as part of the \gls{nn}-based ansatz functions, which is responsible for the strong enforcement of Dirichlet \glspl{bc}.

To overcome limitations related to large variations in the data, normalization has been attempted also in the context of \glspl{pinn} and other neural solvers.
Specific examples include the use of scaling factors \cite{luong2024novel}, hierarchical normalization \cite{le2024hierarchically}, dynamic normalization \cite{deguchi2023dynamic}, normalization by non-dimensionalization \cite{von2024error}, and variable linear transformation \cite{xu2025preprocessing}.
A disadvantage shared by these approaches is that the normalization procedure must be derived anew for each specific problem.
Contrarily, normalization is performed naturally within our method, as it operates on the reference domain of \gls{iga}.

\subsection{Paper organization}
The remaining of this paper is organized as follows. 
In section \ref{sec:v-pinn}, we recall the variational \gls{pinn} framework, which concerns the use of \glspl{nn} to solve energy functional minimization problems arising from variational \gls{pde} formulations.
In section \ref{sec:cad-iga}, we recall some fundamental tools of \gls{cad} and \gls{iga}, to be used throughout the paper.
Section \ref{sec:multipatch-iga-pinn} presents the core methodological contribution of this work, that is, the construction of \gls{nn}-based ansatz functions tailored for multi-patch \gls{cad} domains, as well as the complete neural solver. 
In section \ref{sec:numerical-examples}, our method is applied to two practical use-cases, namely, a 2D magnetostatics simulation of a quadrupole magnet and a 3D solid and contact mechanics simulation of a mechanical holder. 
For each use-case, we outline the theoretical formulation for the particular problem at hand, followed by a comprehensive presentation of the numerical results.
Last, section \ref{sec:conclusion} summarizes the findings of this work and discusses follow-up research possibilities.

\section{Variational physics-informed neural networks}
\label{sec:v-pinn}

We focus on elliptic \glspl{bvp}, for which energy functional formulations exist.
This choice is motivated by the fact that the energy formulation provides a natural optimization setting aligned with \gls{nn} training. 
Additionally, it avoids computing higher order derivatives, which can oftentimes be problematic \cite{sirignano2018dgm}.

As solution function, we assume a scalar-field state, $u: \Omega \rightarrow \mathbb{R}$, where $\Omega \subset \mathbb{R}^d$, $d \in \left\{1,2,3\right\}$, denotes the computational domain.
Typically, there is no general rule for the existence of energy functionals. 
An elliptic differential equation in particular admits an energy functional if and only if it is the Euler–Lagrange equation of a variational problem \cite{evans2022partial}.
This means that the equation can be derived as the condition for a critical point (typically a minimum) of some functional, which usually takes the form
\begin{align}
\label{eq:energy_func}
I\left( u \right) = \int_{\Omega} L\left(\mathbf{x}, u, \textbf{grad}u \right) \, \mathrm{d}\mathbf{x},
\end{align}
where $L$ is a function, typically referred to as the Lagrangian of the associated problem. 
The solution function needs to fulfill the Euler-Lagrange equation, which for a scalar-field state $u$ reads
\begin{align}
\label{eq:euler_lagrange}
\frac{\partial L}{\partial u} - \textrm{div}\left( \frac{\partial L}{\partial \left(\textbf{grad} u\right)}\right) = 0.
\end{align}
Solving the Euler-Lagrange equation \eqref{eq:euler_lagrange} is equivalent to minimizing the energy functional \eqref{eq:energy_func}.
The system's equilibrium state, $u^*$, is obtained as the result of this minimization problem, such that
\begin{equation}
\label{eq:min-problem-u}
u^* = \argmin_{u \in \mathcal{V}} I(u),
\end{equation}
where $\mathcal{V}$ is a function space containing all functions that satisfy the necessary regularity conditions and \glspl{bc} of the \gls{bvp} at hand.
The Lagrangian exists for many elliptic \glspl{pde}, including the use-cases considered in section~\ref{sec:numerical-examples}.
The specific energy functional minimization formulations are derived in sections \ref{sec:num_example_magnetostatics} and \ref{sec:num_example_mechanics}, respectively.
For the same use-cases, $\mathcal{V}$ is the Sobolev space $H^1(\Omega) = \{ u \in L^2 \: : \: \textbf{grad}u \in L^2(\Omega) \}$. 
Accordingly, $u^*$ is the solution to the corresponding \gls{bvp}.

Utilizing the variational \gls{pinn} framework \cite{e2018deep, kharazmi2021hp, khodayi2020varnet}, the trial functions $u \in \mathcal{V}$ are approximated by an \gls{nn}. We denote this approximation as $u_{\bm{\theta}} \approx u$, where $\bm{\theta} \in \Theta$ is a vector containing the \gls{nn}'s trainable parameters and $\Theta$ denotes the corresponding parameter space.
Feedforward \glspl{nn} are used in the following. 
Assuming an \gls{nn} with $L$ layers and $n_l$ neurons in the $l$-th layer, $l=1,\dots, L$, the trainable parameter vector is given as $\bm{\theta} = \left\{\mathbf{W}_l, \mathbf{b}_l\right\}_{l=1}^L$, where $\mathbf{W}_l \in \mathbb{R}^{n_{l-1} \times n_l}$ and $\mathbf{b}_l \in \mathbb{R}^{n_l}$ are the $l$-th layer's weight matrix and bias vector, respectively. For consistency of notation, the input vector has size $n_0$.
Substituting $u$ with $u_{\bm{\theta}}$ in \eqref{eq:min-problem-u}, the \gls{nn} that best approximates the equilibrium state and, accordingly, the solution to the \gls{bvp}, is the one with the parameter vector 
\begin{equation}
\label{eq:min-problem-theta}
\bm{\theta}^* = \argmin_{\bm{\theta} \in \Theta} I\left(u_{\bm{\theta}}\right).
\end{equation}
Note that, in order to have a well-defined problem, the essential \glspl{bc} must be embedded into the structure of the \gls{nn}. That is, $u_{\bm{\theta}}$ must fulfill the essential \glspl{bc} for all $\bm{\theta}\in \Theta$. 
This will be further discussed in section \ref{sec:ansatz_construction}.

From the above, it is obvious that the energy functional is the natural loss function to be minimized during \gls{nn} training, in order to compute the optimal parameter vector $\bm{\theta}^*$.
This is accomplished by first discretizing the energy functional, typically by means of numerical quadrature \cite{e2018deep, sirignano2018dgm, kharazmi2021hp}, and then using a gradient-based optimization method, typically a variant of stochastic gradient descent \cite{bottou1991stochastic, kingma2014method}. The solution of the resulting nonlinear optimization problem yields the set of optimal parameters, i.e., the ones that minimize the discrete energy functional.

\section{Computer-aided design and isogeometric analysis}
\label{sec:cad-iga}

\subsection{Non-uniform rational B-spline}
\label{sec:nurbs}
 
\Gls{nurbs} is the central mathematical tool in \gls{cad}, which provides mathematical representations for modeling curves, surfaces, and volumes \cite{piegl2012nurbs}.
\Gls{nurbs} offers the flexibility and precision needed to represent both simple geometric shapes and complex freeform objects, as it can accurately model both standard analytic forms and arbitrary shapes defined by control points.

The \gls{nurbs} parametrization is defined over a $d$-dimensional reference domain, commonly called the pre-image and here denoted as $\hat{\Omega} = [-1,1]^d$, with its image representing the computational domain $\Omega \subset \mathbb{R}^d$. 
For $\hat{\mathbf{x}} \in \hat{\Omega}$, the \gls{nurbs} parametrization is expressed as
\begin{equation} %
    \bm{f}(\hat{\mathbf{x}}) = \sum\limits_{i_1,\dots,i_d=1}^{n_1,\dots,n_d} \mathbf{p}_{i_1,\dots,i_d} \frac{ w_{i_1,\dots,i_d} b^{(1)}_{i_1} \left( \hat{x}_1 \right) \cdots b^{(d)}_{i_d}(\hat{x}_d)}{\sum\limits_{j_1,\dots,j_d=1}^{n_1,\dots,n_d} w_{j_1,\dots,j_d} b^{(1)}_{j_1}(\hat{x}_1) \cdots b^{(d)}_{j_d}(\hat{x}_d)},
\end{equation}
such that $\bm{f}: \hat{\Omega} \rightarrow \Omega$, where $\mathbf{p}_{i_1,\dots,i_d} \in \Omega$ are control points, $w_{i_1,\dots,i_d}$ are the corresponding weights, and $\left\{b^{(k)}_{i_k}\right\}_{i=1}^{n_k}, k = 1, \dots, d$, is a basis consisting of  univariate B-spline basis functions defined by the Cox–de-Boor formula.
Note that, for non-self-penetrating domains $\Omega$, $\bm{f}$ is a bijective map, equivalently, $\bm{f}^{-1}$ exists and is differentiable.

\subsection{Pushforward and pullback operators}
\label{sec:iga}
The core concept of \gls{iga} is to use \gls{nurbs} as basis functions for the numerical solution of \glspl{pde}, essentially integrating \gls{cad} tools into the \gls{fem}.
Utilizing the \gls{nurbs} parametrization, $\bm{f}$, the solution of the \gls{pde} is first represented on the reference domain and then mapped onto the physical domain. 

Similar to section~\ref{sec:v-pinn}, scalar-field solutions are assumed in both the physical and the reference domain, respectively denoted as $u:\Omega \rightarrow \mathbb{R}$ and $\hat{u}:\hat{\Omega} \rightarrow \mathbb{R}$.  
Then, the \emph{pushforward} operation maps the solution from the reference to the physical domain, and is defined as 
\begin{equation}
\label{eq:pushforward}
u = \bm{f}_* \hat{u} = \hat{u} \circ \bm{f}^{-1}, 
\end{equation}
such that $\bm{f}_*: \hat{\Omega} \rightarrow \Omega$, where the symbol $\circ$ denotes function composition.
The inverse \emph{pullback} operation transforms the solution back to the reference domain, and is defined as 
\begin{equation}
\label{eq:pullback}
\hat{u} = \bm{f}^* u = u \circ \bm{f},
\end{equation}
such that $\bm{f}^*:\Omega \rightarrow \hat{\Omega}$. 

\begin{figure}[t!]
    \centering
    \includegraphics[]{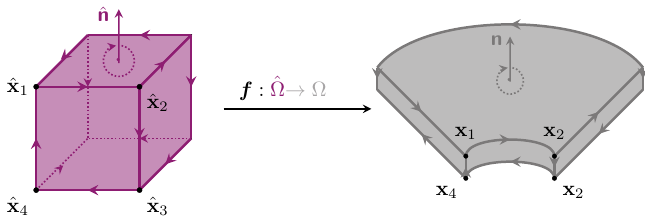}
    \caption{Illustration of the \gls{nurbs}-based geometry map from the reference to the physical domain. We have $\mathbf{x}_k = \bm{f}(\hat{\mathbf{x}}_k)$, $k=1,...,4$. The normal $\hat{\mathbf{n}}$ is transformed to $\mathbf{n}$.}
    \label{fig:iga_geomap}
\end{figure}

Using the integral transformation rule to express the energy functional as an integral over the reference domain and 
integrating the scalar function $u$ in the physical domain leads to 
\begin{align}
\int\limits_{\Omega} u(\mathbf{x}) \text{d}\mathbf{x} = \int\limits_{\hat{\Omega}} \hat{u}(\hat{\mathbf{x}}) \left| \det \left(\partial \bm{f}\right) \right| \text{d}\hat{\mathbf{x}},
\end{align}
where $\partial \bm{f}$ denotes the Jacobian of the \gls{nurbs} parametrization.
The following transformations hold for the gradient and the surface normal:
\begin{align}
\begin{split}
    \mathbf{grad} \: u &= (\partial \bm{f})^{-\top} \mathbf{grad} \: \hat{u}, \\ 
    \mathbf{n} &= |\det (\partial \bm{f})| (\partial \bm{f})^{-\top} \hat{\mathbf{n}},
\end{split}
\end{align}
where $\mathbf{grad} \: \hat{u}$ is the gradient expressed in reference domain coordinates and $\hat{\mathbf{n}}$ is the outer normal on the boundary of the reference domain.  Figure~\ref{fig:iga_geomap} presents an explanatory illustration. 

Note that the pushforward and pullback operations defined in equations \eqref{eq:pushforward} and \eqref{eq:pullback}, respectively, hold for scalar fields only. 
However, in certain cases, they can be applied element-wise to the components of a vector field. 
One such example is considered in the numerical use-case presented in section \ref{sec:num_example_mechanics}.

\subsection{Multi-patch CAD geometries}
\label{sec:multipatch_iga}
A single \gls{nurbs} has only limited ability to represent a wide variety of shapes, thus making it unsuitable for parametrizing arbitrary, non-trivial \gls{cad} domains. Consequently, multiple \gls{nurbs} parametrizations are required, each parametrizing  a subdomain of the geometry. Collectively, these subdomains form the physical geometry.
In the following, the term ``patch'' is used interchangeably for both a subdomain of the physical geometry and its corresponding \gls{nurbs} parametrization.

We assume $N_{\Omega}$ \gls{nurbs} parametrizations, denoted as $\bm{f}_{i}$, $i=1, \ldots, N_{\Omega}$, each representing a subdomain $\Omega_i \subset \Omega$. 
To construct a multi-patch geometry, the individual patches are generated by mapping a reference domain $\hat{\Omega}_i = \left[-1,1\right]^d$ onto the physical geometry, such that $\bm{f}_{i}\left(\hat{\Omega}_i\right) = \Omega_i$. 
This mapping ensures that the collection $\left\{\Omega_i\right\}_{i=1}^{N_\Omega}$ forms a domain decomposition, such that $\Omega = \bigcup_{i=1}^{N_\Omega} \Omega_i$.
Note that, although the reference domains $\hat{\Omega} := \hat{\Omega}_1 = \cdots = \hat{\Omega}_{N_{\Omega}} = [-1,1]^d$ are identical, individual indexing is necessary, because the points $\hat{\mathbf{x}} \in \hat{\Omega}_i$ are mapped differently onto $\Omega$, depending on the patch-specific \gls{nurbs} parametrization $\bm{f}_{i}$.

In this work, the patches are assumed to be non-overlapping and conforming at their interfaces. Non-overlapping means that the interiors of the patches are mutually disjoint. 
Denoting the boundary of a subdomain $\Omega_i$ as $\partial\Omega_i$, this means that $(\Omega_i \setminus \partial \Omega_i) \cap (\Omega_j \setminus \partial \Omega_j) = \emptyset$, for all $i, j = 1, \ldots, N_\Omega$. 
Conformity implies that, for any two adjacent patches $\Omega_{i}$ and $\Omega_{j}$, $i \neq j$, with a nonempty $(d-1)$-dimensional interface $\Gamma_{ij} := \partial \Omega_i \cap \partial \Omega_j$, the pre-images $\hat{\Gamma}_{i\rightarrow j} = \bm{f}^{-1}_{i}(\Gamma_{ij})$ and $\hat{\Gamma}_{j \rightarrow i}= \bm{f}^{-1}_{j}(\Gamma_{ij})$ must correspond to facets of the respective reference domains $\hat{\Omega}_i$ and $\hat{\Omega}_j$. 
Additionally, the tangent vectors at any point $\mathbf{x} \in \Gamma_{ij}$, computed using the two different parametrizations, must coincide \cite{kleiss2012ieti}.

\begin{figure}[t!]
    \centering
    \begin{tikzpicture}

     \definecolor{darkblue}{rgb}{0.0,0.0,0.9}
     \definecolor{darkred}{rgb}{0.9,0.0,0.0}
     \definecolor{darkgreen}{rgb}{0.0,0.75,0.0}

     \draw[very thick] (0,-1) rectangle (2,1);
     \node at (1,0.0) { $\hat{\Omega}$};
     \begin{scope}[shift={(0.2,-0.8)}]
         \draw[darkblue,->,-stealth, dash dot, thick] (0,0) -- (1.6,0) node[anchor=south] {$\mathbf{\hat{x}}_1$};
         \draw[darkred,->,-stealth,dashed, thick] (0,0) -- (0,1.6) node[anchor=west] {$\mathbf{\hat{x}}_2$};
     \end{scope}

     \begin{scope}[xshift=3cm]
\pgftransformnonlinear{\mytransformation}
         \draw[very thick] (1,-2) rectangle (3,0);
         \draw[very thick] (1,0) rectangle (3,2);
         \draw[very thick] (3,-2) rectangle (5,0);
         \draw [TUDa-10c, line width = 2] (1,-2) -- (1,2) -- (3,2);
         \node at (2,-1) {$\Omega_2$};
         \node at (2,1) {$\Omega_1$};
         \node at (4,-1) {$\Omega_3$};
        \node[TUDa-10c] at (0.4,2) {$\Gamma_{\mathrm{D}}$};

         \begin{scope}[shift={(1.2,-1.8)}]
             \draw[darkblue,->,-stealth, dash dot, thick] (0,0) -- (1.6,0) node[anchor=south east] {};
             \draw[darkred,->,-stealth, dashed, thick] (0,0) -- (0,1.6) node[anchor=west] {};
         \end{scope}

         \begin{scope}[shift={(1.2,0.2)}]
             \draw[darkblue,->,-stealth, dash dot, thick] (0,0) -- (1.6,0) node[anchor=south] {};
             \draw[darkred,->,-stealth, dashed, thick] (0,0) -- (0,1.6) node[anchor=west] {};
         \end{scope}

         \begin{scope}[shift={(3.2,-1.8)}]
             \draw[darkblue,->,-stealth, dash dot, thick] (0,0) -- (1.6,0) node[anchor=south east] {};
             \draw[darkred,->,-stealth,dashed, thick] (0,0) -- (0,1.6) node[anchor=west] {};
         \end{scope}
 
         \draw[red] (3,0) node[] {\Large \textbullet};
         \node [anchor=south west, red] at (3,0.25) {$\boldsymbol{\xi}_{123}$};

     \end{scope}

     \draw[->, -stealth] (1,1.0) .. controls (2.5,1.75) .. (4.3,1.5) node[midway, above] {$\bm{f}_{1}$};;
     \draw[->, -stealth] (2,0.0) .. controls (3,-0.5) .. (4.1,-0.5) node[midway, above right] {$\bm{f}_{2}$};;
     \draw[->, -stealth] (1,-1.0) .. controls (2,-3.5) and (8.0,-2.5)  .. (8.0,-1.3) node[midway, above] {$\bm{f}_{3}$};;
\end{tikzpicture}
    \caption{Example of a two-dimensional domain represented by three patches, $\Omega_1$, $\Omega_2$, $\Omega_3$. The reference domain $\hat{\Omega}$ and the mapping of the reference coordinate system onto each patch are also shown (red and blue dotted lines). 
    The Dirichlet boundary $\Gamma_{\mathrm{D}}$ is marked in purple. The point $\bm{\xi}_{123} \in \Omega$ belongs to all three patches simultaneously.}
    \label{fig:multipatch_example}
\end{figure}
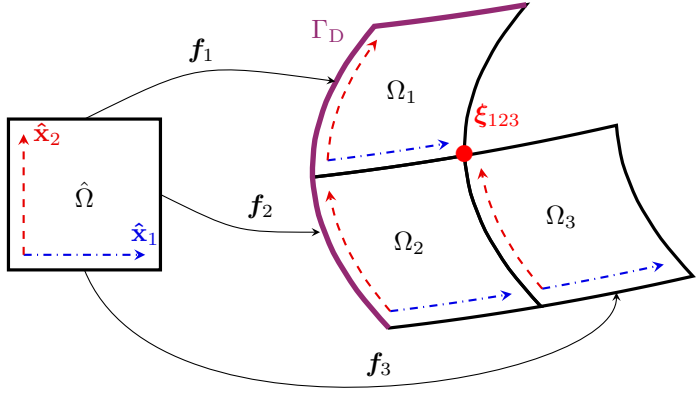

Figure \ref{fig:multipatch_example} illustrates an example of three conforming patches on a generic two-dimensional domain. The reference domain $\hat{\Omega}$ is mapped onto the three patches $\Omega_i$, $i=1, 2, 3$, where conformity is enforced at their interfaces and on points where multiple subdomains meet. Each interface between the individual patches corresponds to either an edge or a corner in the respective reference domain.

\section{Multi-patch isogeometric neural solver}
\label{sec:multipatch-iga-pinn}

\subsection{Neural network-based ansatz in the reference domain}
\label{sec:ansatz_construction}

The main building blocks of the suggested neural solver are \gls{nn}-based ansatz functions capable of approximating solutions within the reference domain of \gls{iga}, to be then transformed into solutions on the physical geometry. 
In the following we assume that, for each subdomain (patch) $\Omega_i$, the solution in the reference domain is represented by \gls{nn}-based ansatz functions $\hat{u}_{\bm{\theta}_i} : \hat{\Omega}_i \rightarrow \mathbb{R}$, $i=1,\dots, N_{\Omega}$.
These ansatz functions consist of products and superpositions of \glspl{nn} and polynomials, defined on the reference domain.
To ease the notation, the vector $\bm{\theta} \in \Theta$ concatenates the trainable parameters of all \glspl{nn} employed in the approximation, such that $\bm{\theta} = \left\{\bm{\theta}_i\right\}_{i=1}^{N_{\Omega}}$. 
Accordingly, the patch-specific \gls{nn} $\hat{u}_{\bm{\theta}_i}$ contains the trainable parameters that correspond to the $i$-th patch only.

The construction of the ansatz functions $\left\{\hat{u}_{\bm{\theta}_i}\right\}_{i=1}^{N_\Omega}$ must satisfy the Dirichlet \glspl{bc}, while also ensuring differentiability within each subdomain and $C^0$-continuity across the interfaces between patches. 
These conditions are particularly advantageous when addressing discontinuous material properties between patches. 
Utilizing the previously defined pushforward and pullback operations, continuity requires that 
\begin{align}
\label{eq:continuity_cond}
\hat{u}_{\bm{\theta}_i} \left(\hat{\Gamma}_{i \rightarrow j}\right) = \hat{u}_{\bm{\theta}_j}\left(\hat{\Gamma}_{j \rightarrow i}\right),
\end{align}
for every interface $\Gamma_{ij}$.
Similar to classical approaches found in \gls{iga}, condition \eqref{eq:continuity_cond} ensures a continuous transition between individual patches of the multi-patch geometry, by constraining the solution representation in the reference domain.
In order to correctly represent situations where boundary elements of different dimensionality intersect, it is necessary to introduce multi-indices. 
This necessity arises in particular at points where three or more patches of a multi-patch geometry coincide. To guarantee that the continuity condition \eqref{eq:continuity_cond} is satisfied in such configurations, the interface contributions must enforce continuity simultaneously across all interacting patches.
To fulfill these requirements, we define our ansatz function within each patch as a superposition of an interior term, $\hat{u}_{\bm{\hat{\theta}}_i}^{\text{int}}$, representing the solution inside each patch and interface-conforming terms, $\hat{u}_{\bm{\hat{\theta}}_{\bm{j}}}^{\bm{j} \rightarrow i}$, imposing continuity across neighboring patches.
The definition reads as follows:
\begin{align}
\hat{u}_{\bm{\theta}_i}^{(i)} = \hat{u}_{\bm{\hat{\theta}}_i}^{(i,\text{int})} + \sum_{\bm{j} \in \mathcal{J}(i)}\hat{u}_{\bm{\hat{\theta}}_{\bm{j}}}^{\bm{j} \rightarrow i}, 
\end{align}
where $\mathcal{J}(i)$ is  a multi-index set associated with the $i$-the patch, which comprises the multi-indices of all topological entities where the $i$-th patch intersects with other patches, i.e., surfaces (2D), curves (1D), and points (0D), and $\bm{j}$ are the multi-indices contained therein.
For example, considering the two-dimensional domain depicted in Figure~\ref{fig:multipatch_example}, $\mathcal{J}(1) = \left\{(1,2), (1,2,3)\right\}$ due to the interface $\Gamma_{12}$ shared by subdomains $\Omega_1$ and $\Omega_2$ and the point $\xi_{123}$ shared by all three subdomains.
Accordingly, $\mathcal{J}(2) = \left\{(1,2), (2,3), (1,2,3)\right\}$ and $\mathcal{J}(3) = \left\{(2,3), (1,2,3)\right\}$.
The interface terms $\hat{u}_{\bm{\hat{\theta}}_{\bm{j}}}^{\bm{j} \rightarrow i}$, $\bm{j}\in\mathcal{J}(i)$, connect intersecting patches.

The term $ \hat{u}_{\bm{\hat{\theta}}_i}^{(i,\text{int})}$ contributes solely  to the interior of the subdomain $\Omega_i$ and vanishes on the interfaces defined by the multi-index set $\mathcal{J}(i)$. 
The trainable parameters per patch are then given as $\bm{\theta}_i = \bm{\hat{\theta}}_i \cup \bigcup_{\bm{j} \in \mathcal{J}(i)}\bm{\hat{\theta}}_{\bm{j}} $.
The ansatz functions $\hat{u}_{\bm{\theta}_i}$ are evaluated separately for the specific patch they correspond to, and collectively yield a global solution.
The corresponding ansatz space $\mathcal{V}_{\Theta}$ for the physical domain is then defined by the individual parameter spaces $\Theta$ of the \glspl{nn} and the patch-wise inverse pullback:
\begin{align}
\mathcal{V}_{\Theta} := \left\{ \hat{u}_{\bm{\theta}_i} \circ \bm{f}_{i}^{-1}\,:\, i=1,...,N_\Omega \right\}.
\end{align}
The parameter space $\Theta$ corresponds to the  Cartesian product of the patch-specific parameter spaces which include the parameters of the \glspl{nn} $\hat{u}_{\bm{\theta}_i}^{(i)}$, $i=1, \dots, N_{\Omega}$.
The global ansatz function (in the physical domain) is denoted as $u_{\bm{\theta}} \in \mathcal{V}_{\Theta}$.
Note that, for vector-valued ansatz functions $\hat{\bm{u}}: \hat{\Omega} \rightarrow \mathbb{R}^n$, the considerations outlined above can be applied independently to each component (see section \ref{sec:num_example_mechanics}).

\subsubsection{Ansatz inside the domains}
\label{sec:ansatz-in-domain}

We first focus on constructing the ansatz within the domain. 
We assume that the Dirichlet boundary, $\Gamma_\mathrm{D}$, is derived by applying the pushforward map onto the respective facets of the reference domains.
Further, we assume that the Dirichlet data is defined by a function $g_\mathrm{D}:\Gamma_\mathrm{D} \rightarrow \mathbb{R}^d$ on the Dirichlet boundary.
The ansatz is composed of the product of the output of the \gls{nn} used to approximate the solution in the reference domain, and a polynomial that ensures the enforcement of Dirichlet boundary conditions. 
The value of the ansatz on the Dirichlet boundary is imposed by adding an additional term to the ansatz.   
Specifically, the ansatz inside the domain is defined as
\begin{align}
    \hat{u}^{(i,\text{int})}_{\hat{\bm{\theta}}_i}\left(\hat{\mathbf{x}}\right) = \mathcal{N}_{\hat{\bm{\theta}}_i}\left(\hat{\mathbf{x}}\right)  \hat{p}(\hat{\mathbf{x}}) + (g_\mathrm{D} \circ \bm{f})(\hat{\mathbf{x}}) \hat{p}_\mathrm{D}(\hat{\mathbf{x}}),
\end{align}
where $\mathcal{N}_{\hat{\bm{\theta}}_i} : [-1,1]^d \rightarrow\mathbb{R}$ is an \gls{nn} depending on the trainable parameters $\hat{\bm{\theta}}_i$ and the polynomial $\hat{p}(\hat{\mathbf{x}}) = \prod_{k=1}^{d} (\hat{x}_k-1)^{\alpha_k}(\hat{x}_k+1)^{\beta_k}$ is chosen so that its pullback vanishes on interfaces and Dirichlet boundaries i.e.
\begin{align}
    \hat{p} \circ \bm{f}^{-1}_{i} = 0\quad \text{for}\quad \mathbf{x}\in \partial \Omega_i \cap \left( \Gamma_\mathrm{D} \cup \bigcup\limits_{i\neq j} \partial \Omega_j \right).\label{eq:set_zeros}
\end{align}
This construction ensures that the product of $\mathcal{N}_{\hat{\bm{\theta}}_i}$ and $\hat{p}$ yields a parametric function which does not influence Dirichlet boundaries and patch interfaces, however, is variable inside the patch interiors.
The inclusion of interfaces in \eqref{eq:set_zeros} is intentional, as the interface is handled separately by an additional term introduced in the next section.
Using similar principles, we construct the term representing the Dirichlet boundary condition.
The choice of the polynomial $\hat{p}_\mathrm{D}$ must ensure that 
\begin{align}
\begin{split}
\hat{p}_\mathrm{D} \circ \bm{f}^{-1}_{i} &= 1 \text{ for }\mathbf{x} \in \Gamma_\mathrm{D},\\
\hat{p}_\mathrm{D} \circ \bm{f}^{-1}_{i} &= 0 \text{ for }\mathbf{x} \in \Gamma^{\text{opp}}_\mathrm{D},
\end{split}
\end{align}
where $\Gamma^{\text{opp}}_\mathrm{D}$ denotes the interface opposing $\Gamma_\mathrm{D}$.
From an implementation point of view, this approach can be interpreted as a custom output layer with a residual connection.

As an example, we revisit the 2D domain composed by 3 patches shown in Figure \ref{fig:multipatch_example}. 
In the same figure, the mapping of the reference coordinate system onto the physical subdomains is also shown. 
For simplicity, we assume homogeneous Dirichlet boundary conditions i.e. $g_D = 0$ on $\Gamma_\mathrm{D}$, where $\Gamma_{\mathrm{D}}$ is represented by the green line.  
In the reference domain of $\Omega_2$, this translates to $\hat{x}_1=-1$, while for $\Omega_2$ it maps to $\hat{x}_2=1 \text{ and } \hat{x}_1=-1$. 
There are two interfaces, $\Gamma_{12}$ and $\Gamma_{13}$, and all three domains share a common point $\bm{\xi}_{123}$ for the multi-index $\bm{j} = \left(1,2,3\right)$.
Then, the boundary-conforming ansatz inside the domains are:
\begin{align}
\begin{split}
    \hat{u}^{(1, \text{int})}_{\hat{\bm{\theta}}_1}(\hat{\mathbf{x}}) &= \mathcal{N}_{\hat{\bm{\theta}}_1}(\hat{\mathbf{x}})(\hat{x}_1+1)(\hat{x}_2+1)(\hat{x}_2-1),\\
    \hat{u}^{(2, \text{int})}_{\hat{\bm{\theta}}_2}(\hat{\mathbf{x}}) &= \mathcal{N}_{\hat{\bm{\theta}}_2}(\hat{\mathbf{x}})(\hat{x}_1+1)(\hat{x}_1-1)(\hat{x}_2-1),\\
    \hat{u}^{(3, \text{int})}_{\hat{\bm{\theta}}_3}(\hat{\mathbf{x}}) &= \mathcal{N}_{\hat{\bm{\theta}}_3}(\hat{\mathbf{x}})(\hat{x}_1+1)(\hat{x}_2+1)(\hat{x}_2-1).
\end{split}
\end{align}

\subsubsection{Ansatz on domain interfaces}
\label{sec:ansatz-interfaces}

To connect the solutions defined on individual patches, we construct interface-conforming terms. This process is iterative and can be applied to an arbitrary number of dimensions.

We first consider the 0-dimensional intersections that do not belong to the Dirichlet boundary:
\begin{align}
    \Xi_0 = \Big\{ \partial\Omega_i \cap \partial\Omega_j \; : \; \partial\Omega_i \cap \partial\Omega_j \text{ is 0-dimensional and } \partial\Omega_i \cap \partial\Omega_j \nsubseteq \Gamma_\mathrm{D} \Big\}.
\end{align}
Note that the indices from the set $\Xi_0$ may have different sizes. Since an element of $\Xi_0$ may be shared by multiple patches, its pre-image corresponds to different points in the respective reference domains. 
Given a point $\bm{\xi}\in\Xi_0$, let $\bm{j} = \mathcal{I}(\bm{\xi})$ contain the indices of all patches sharing the 0-dimensional interface $\bm{\xi}$. Then, the ansatz in all the patches $i\in \bm{j}$ is expressed as
\begin{align}
    \hat{u}_{\hat{\bm{\theta}}_{\bm{j}}}^{\bm{j} \rightarrow i}(\hat{\mathbf{x}}) = \hat{\bm{\theta}}_{\bm{j}} \prod_{k=1}^d \bigl( (\hat{x}_k-1)^{\alpha_k} (\hat{x}_k+1)^{\beta_k} \bigr), \label{eq:single_param}
\end{align}
where the exponents $\alpha_k$ and $\beta_k$ are chosen so that the function vanishes at all vertices of the reference domain but $\bm{f}^{-1}_{i}(\bm{\xi})$. 
In this case, a single trainable parameters $\hat{\bm{\theta}}_{\bm{j}}$ are responsible for the value of the ansatz on $\bm{\xi}$.

Next, we construct the $q$-dimensional interface ansatz based on the sets $\Xi_{q-1}, \dots, \Xi_0$. The set $\Xi_q$ contains all the $q$-dimensional interfaces:
\begin{align}
    \Xi_q = \Big\{ \partial \Omega_i \cap \partial \Omega_j \; : \; \partial \Omega_i \cap \partial \Omega_j \text{ is } q\text{-dimensional and } \partial\Omega_i \cap \partial\Omega_j \nsubseteq \Gamma_\mathrm{D} \Big\}.
\end{align}
Similarly to the 0-dimensional case, for an element $\bm{\xi} \in \Xi_q$ shared by patches indexed by $\bm{j}=\mathcal{I}(\bm{\xi})$, the corresponding ansatz in the $i$-th patch, $i\in\bm{j}=\mathcal{I}(\bm{\xi})$, is defined by
\begin{align}
    \hat{u}_{\hat{\bm{\theta}}_{\bm{j}}}^{\bm{j} \rightarrow i}(\hat{\mathbf{x}}) = \mathcal{N}_{\hat{\bm{\theta}}_{\bm{j}}}(\hat{x}_{l_1}, \dots, \hat{x}_{l_q}) \prod_{k=1}^d \bigl( (\hat{x}_k-1)^{\alpha_k} (\hat{x}_k+1)^{\beta_k} \bigr),
\end{align}
where the \gls{nn} $\mathcal{N}_{\hat{\bm{\theta}}_{\bm{j}}}$ is defined on the pre-image $\bm{f}^{-1}_{i}(\bm{\xi})$, and the coefficients $\alpha_k$ and $\beta_k$ are chosen such that 
$ \hat{u}_{\hat{\bm{\theta}}_{\bm{j}}}^{\bm{j} \rightarrow i}(\hat{\mathbf{x}}) \circ \bm{f}^{-1}_{i}$
vanishes on all $q$-dimensional faces of the reference domain that do not contain $\bm{\xi}$, as well as on all elements of $\Xi_{q-1}\cup \cdots \cup \Xi_0$. This latter condition ensures that the contribution from the $q$-dimensional interface does not overlap with those from lower-dimensional interfaces.
The procedure is iteratively repeated until the set $\Xi_{d-1}$ and the corresponding ansatz functions are constructed.

As an example, we revisit the domain presented in Figure \ref{fig:multipatch_example}. 
In this case, the set $\Xi_0=\{\bm{\xi}_{123}\}$ contains a single element and its corresponding index set is $\mathcal{I}(\bm{\xi}_{123})=(1,2,3)$. 
Then, the interface ansatz functions are
\begin{align}
\label{eq:many_domains}
\begin{split}
    \hat{u}^{(1,2,3) \rightarrow 1}&=\hat{\bm{\theta}}_{(1,2,3)}(\hat{x}_1+1)(\hat{x}_2-1), \\
    \hat{u}^{(1,2,3) \rightarrow 2}&=\hat{\bm{\theta}}_{(1,2,3)}(\hat{x}_1+1)(\hat{x}_2+1), \\
    \hat{u}^{(1,2,3) \rightarrow 3}&=\hat{\bm{\theta}}_{(1,2,3)}(\hat{x}_1-1)(\hat{x}_2+1),
\end{split}
\end{align}
where $\hat{\bm{\theta}}_{(1,2,3)}$ are the trainable parameters corresponding to the common point $\bm{\xi}_{123}$. The set $\Xi_1$ contains the two interfaces $\Gamma_{12}$ and $\Gamma_{23}$, with $\mathcal{I}(\Gamma_{12})=(1,2)$, $\mathcal{I}(\Gamma_{23})=(2,3)$. The corresponding interface functions are
\begin{align}
\begin{split}
    \hat{u}^{(1,2) \rightarrow 1}_{\hat{\bm{\theta}}_{(1,2)}}&=\mathcal{N}_{\hat{\bm{\theta}}_{(1,2)}}(\hat{x}_1)(\hat{x}_1+1)(\hat{x}_1-1)(\hat{x}_2-1), \\
    \hat{u}^{(1,2) \rightarrow 2}_{\hat{\bm{\theta}}_{(1,2)}}&=\mathcal{N}_{\hat{\bm{\theta}}_{(1,2)}}(\hat{x}_1)(\hat{x}_1+1)(\hat{x}_1-1)(\hat{x}_2+1), \\
    \hat{u}^{(2,3) \rightarrow 2}_{\hat{\bm{\theta}}_{(2,3)}}&=\mathcal{N}_{\hat{\bm{\theta}}_{(2,3)}}(\hat{x}_2)(\hat{x}_1+1)(\hat{x}_2+1)(\hat{x}_2-1), \\
    \hat{u}^{(2,3) \rightarrow 3}_{\hat{\bm{\theta}}_{(2,3)}}&=\mathcal{N}_{\hat{\bm{\theta}}_{(2,3)}}(\hat{x}_2)(\hat{x}_1-1)(\hat{x}_2+1)(\hat{x}_2-1).
\end{split}
\end{align}
In this case, the \glspl{nn} used are restricted to only one of the coordinates. 
Moreover, the Dirichlet boundary as well as the set $\Xi_0$ are taken into account for the interfaces corresponding to $\Gamma_{12}$.

\begin{figure}[t!]
\centering
\includegraphics[width = 0.495\linewidth]{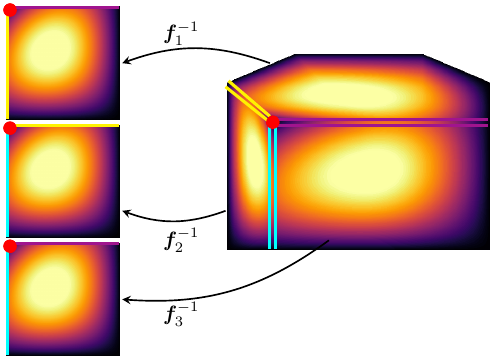}
\includegraphics[width = 0.495\linewidth]{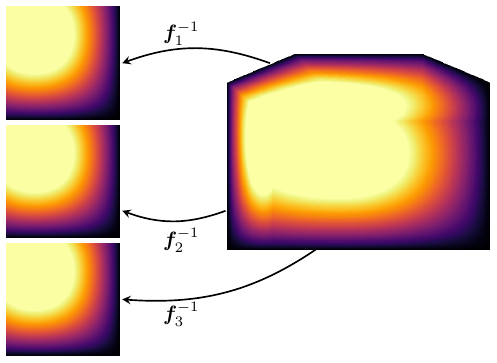}
\caption[]{
Illustration of the construction of interface ansatz functions. At the point where multiple subdomains meet, indicated by the red circle, the interface functions take zero values. 
Continuity is enforced at the interfaces indicated by the colored lines (yellow, cyan, and purple) in the reference and physical domains.
}
\label{fig:a_coin}
\end{figure}

To illustrate how the ansatz functions are constructed on domain interfaces, we refer to Figure \ref{fig:a_coin}. 
On the left-hand side, the interface functions from equation \eqref{eq:many_domains} spanning three subdomains are omitted, but the planes where continuity must be enforced in both the parametric and physical domains are indicated. 
The red circle marks the coincidence point of all three subdomains (similar to point $\bm{\xi}_{123}$ in Figure \ref{fig:iga_geomap}).
On the right-hand side, the complete ansatz function is shown, incorporating all interface functions. 
As can be observed, continuity is enforced across all patches.

\section{Numerical examples}
\label{sec:numerical-examples}

In this section, the suggested neural solver is used for, first, a 2D magnetostatics simulation, and, second, a 3D solid mechanics simulation involving nonlinear material behavior and nonlinear \glspl{bc}. 
In both examples, we begin with the formulation of the problem at hand and then proceed with implementation details and the corresponding numerical results.

\subsection{Magnetostatic simulation of a quadrupole magnet}
\label{sec:num_example_magnetostatics}

\subsubsection{Magnetostatic formulation}
\label{sec:static_bvp}
We consider the magnetostatic problem setting, which can be retrieved by considering the eddy-current problem and dropping the time dependencies and conductive materials \cite{jackson2021classical}.
Under these assumptions, the problem can be formulated in terms of the magnetic vector potential. If the problem is reduced to 2D, only the longitudinal component $u: \Omega \rightarrow \mathbb{R}$ of the magnetic vector potential is required \cite{jackson2021classical}. In this case, the governing equation for a linear isotropic material is
\begin{align}
\label{eq:magnetostatic_strong}
\begin{split}
  - \mathbf{grad} \left(\nu \: \mathbf{grad} u \right)& = j_z, \quad \text{in } \Omega, \\
  u & = 0, \quad \text{on } \Gamma_\mathrm{D}, \\
  \nu \: \mathbf{grad} u \cdot \bm{n} & = 0, \quad \text{on } \Gamma_N,
  \end{split}
\end{align}
where $\Omega$ is the computational domain, $j_z$ is the $z$-component of the excitation current density, $\GD$ the Dirichlet boundary and $\GN$ the Neumann boundary.
An energy functional minimization problem of the form \eqref{eq:min-problem-u} can be derived, where the energy functional is given as
\begin{align}
    I(u) = \frac{1}{2}\int\limits_{\Omega} \nu\: \mathbf{grad} u \cdot \mathbf{grad} u \text{ d}\mathbf{x} - \int\limits_{\Omega} u j_z \text{ d} \mathbf{x}.
\end{align}
By taking into consideration a multi-patch decomposition of the domain $\Omega$ as shown in section \ref{sec:multipatch_iga}, the evaluation of the energy functional in the reference domain yields
\begin{align}
\label{eq:en_funct}
I\left(\hat{u}_{\bm{\theta}} \right)=\frac{1}{2} \sum_{i=1}^{N_{\Omega}} \int_{\hat{\Omega}_i} \nu \, 
\textbf{grad} \left( \hat{u}_{\bm{\theta}} \right)^{\intercal} \,\mathbf{K}^{(i)}(\hat{\mathbf{x}}) \,\textbf{grad} \left( \hat{u}_{\bm{\theta}} \right)\,
\mathrm{d}\hat{\mathbf{x}}-\int_{\hat{\Omega}_{\text{s}}} {j}_z \cdot \hat{u}_{\bm{\theta}} \left|\det D \bm{f}_{i} \right| \, \mathrm{d} \hat{\mathbf{x}},
\end{align}
where $\mathbf{K}^{(i)}(\hat{\mathbf{x}})=(D \bm{f}_{i})^{-\intercal} (D \bm{f}_{i})^{-1}\left|\det D \bm{f}_{i} \right|$ and $\hat{\Omega}_{\text{s}}$ is the reference domain corresponding to the patch that contains the source current excitation. 
The ansatz function $\hat{u}_{\bm{\theta}}$ is constructed to naturally enforce homogeneous Dirichlet boundary conditions, hence, we do not require additional penalization terms.

For the material distribution, we consider linear magnetic materials, where $\mu_0 = 4\pi \cdot 10^{-7} \si{\frac{H}{m}}$ is the permeability of free space, $\mu_{\text{Fe}}=2000 \mu_0$ the permeability of iron, and $\mu_{\text{Cu}}\approx \mu_0$ the permeability of copper.
For the excitation, we consider a constant current density ${j}_z=7.95\cdot10^{8} \si{\frac{A}{m^2}}$.

To quantify error, we use the relative $L^2$ error, given by the formula
\begin{align}
\label{eq:rel-L2-error}
\epsilon_{\text{rel}, L^2(\Omega)} = \frac{\left\|u_{\bm{\theta}} - u^\text{\gls{fem}} \right\|_{L^2(\Omega)}}{\left\|u^\text{\gls{fem}} \right\|_{L^2(\Omega)}},
\end{align}
where $u^\text{\gls{fem}}$ is a highly resolved \gls{fem} solution.
For both computational examples, we calculate the relative $L^2$ error, once on the complete domain $\Omega$ and for the individual subdomains $\Omega_1,...,\Omega_{N_{\Omega}}$.
This approach enables us to assess the overall error across the domain and analyze its distribution.

\subsubsection{Simple quadrupole magnet model}
\label{subsec:num_prob1}

\begin{figure}[t!]
\centering
\includegraphics[width = 0.45\linewidth]{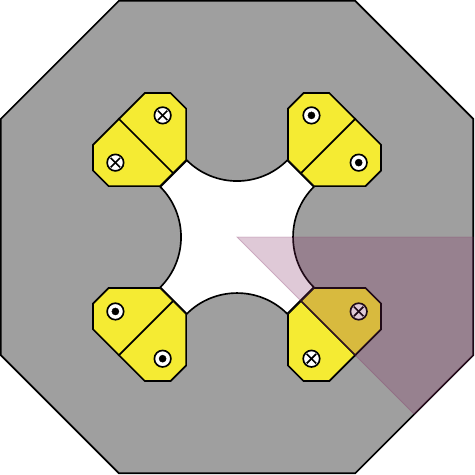}
\includegraphics[width = 0.45\linewidth]{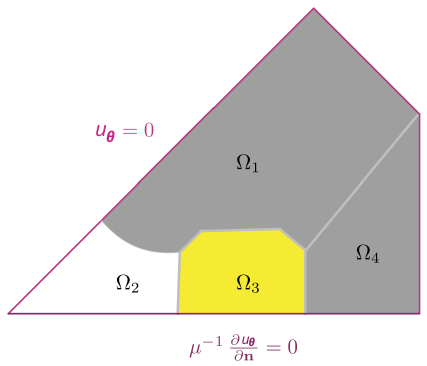}
\caption[Simple quadrupole model geometry]{Domain geometry of the simple quadrupole model. The iron yoke is shown in gray, the current excitations in yellow, and the air gap in white. The purple shading indicates the model section in the symmetry planes. \textbf{Left:} Full cross-section. 
\textbf{Right:} One-eighth of cross-section, exploiting rotational symmetry. The domain is partitioned into patches $\Omega_i$, $i=1, \dots,4$.}
\label{fig:geo_simple_quad}
\end{figure}

First, we consider a relatively simple quadrupole magnet geometry, depicted in Figure \ref{fig:geo_simple_quad}.
On the left hand side of Figure \ref{fig:geo_simple_quad}, the complete cross-section of the quadrupole magnet is depicted. 
On the right hand side of Figure \ref{fig:geo_simple_quad}, one eighth of the magnet's geometry is shown, exploiting its rotational symmetry. 
The latter constitutes the domain geometry of the computational model and is partitioned into patches $\Omega_1, \dots,\Omega_4$.
The characterization ``simple'' is due to the fact that only four patches are required for the complete parametrization of $\Omega$.
The purple shading highlights the location of the computational model's domain on the complete cross-section domain.
The iron yoke is shown in gray, the current excitations in yellow, and the air domain in white.
Dirichlet and Neumann boundary conditions are applied at the boundaries as indicated.

A \gls{resnet} architecture \cite{he2016deep} is chosen for the \glspl{nn} on both the subdomains and on the interfaces.
In particular, we employ \glspl{nn} with four \gls{fc} layers, including an input layer, one \gls{resnet} block, and an output layer.
The number of neurons in each layer varies depending on whether the \gls{nn} is allocated to the domain or to an interface. That is, more neurons are allocated for the domain compared to the interface.
For the domain-\glspl{nn}, we employ $16$ neurons per layer, and the input and output dimensions for each \gls{fc} layer vary throughout the network according to $(2,16)\rightarrow(16,16)\rightarrow(16,16)\rightarrow(16,1)$.
For the interface-\glspl{nn}, we employ eight neurons per layer, and the input and output dimensions for each \gls{fc} layer vary according to $(1,8)\rightarrow(8,8)\rightarrow(8,8)\rightarrow(8,1)$.
In both cases, there exist two skip connections, before the second and fourth layer, respectively.
The hyperbolic tangent activation function is used.
The \glspl{nn} are trained for 1000 epochs, utilizing $ 4 \times 10^4 $ Monte Carlo sampling points in each epoch. 
For the reference solution, we employ the FEniCS \gls{fem} solver \cite{AlnaesEtal2014, A.Logg.2010}, using a first-order tetrahedral mesh consisting of  $13957$ elements. %
Table \ref{table:train_quad_simple} provides an overview of the \gls{nn} and trainable parameters employed in this use-case. 

\begin{table}[b!]
	\centering	\caption{Distribution of \glspl{nn} and trainable parameters for the simple quadrupole model.}
	\begin{tabular}{l c c}
		\hline\hline
		  & Number of \glspl{nn} &  Number of trainable parameters \\ [0.5ex] 
		\hline
		Point interfaces (0D)				& -- &	2 		 \\[0.5ex] 
		Interface \glspl{nn} (1D) 					&5&	108 		 \\[0.5ex] 
		Domain \glspl{nn} (2D) 				&4&	360		 \\[0.5ex] 
		Total 				& 9 &	1982	 	 \\[0.5ex] 
		\hline
	\end{tabular}
	\label{table:train_quad_simple}
\end{table}

\begin{figure}[t!]
\centering
\includegraphics[width = 0.8\linewidth]{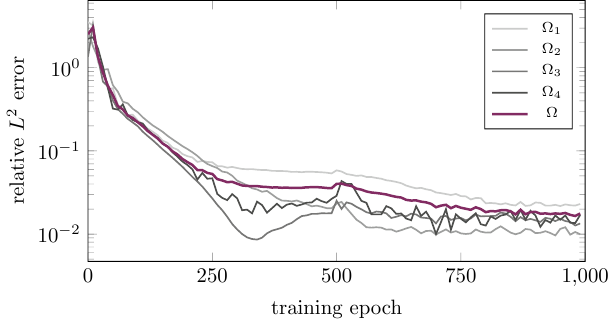}
\caption{Relative $L^2$ error during training for the entire computational domain and for each subdomain and the simple quadrupole model.}
\label{fig:error_simple_quadrupole}
\end{figure}

\begin{figure}[t!]
\centering
\begin{subfigure}{0.49\textwidth}
    \centering
    \includegraphics[width=\linewidth]{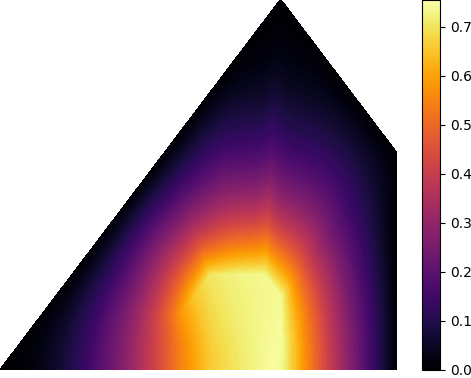}
    \subcaption{\gls{pinn} solution.}
    \label{fig:simple_model_solution}
\end{subfigure}
\hfill
\begin{subfigure}{0.49\textwidth}
    \centering
    \includegraphics[width=\linewidth]{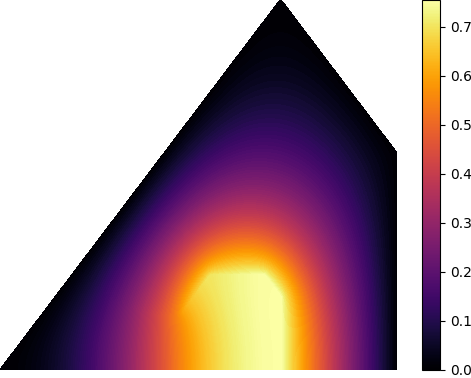}
        \subcaption{Reference \gls{fem} solution.}
        \label{fig:simple_reference}
\end{subfigure}
\begin{subfigure}{0.49\textwidth}
    \centering
    \includegraphics[width=\linewidth]{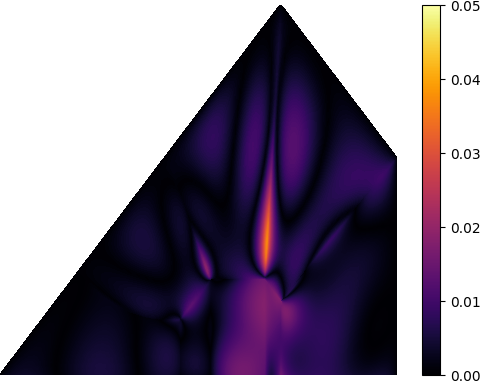}
    \subcaption{Pointwise error.}
    \label{fig:simple_absolute_error}
\end{subfigure} 
\caption{Field solutions and corresponding error field for the simple quadrupole magnet model.}
\label{fig:contour_simple_quadrupole}
\end{figure}

A plot of the relative $L^2$ error over the entire training process is displayed in Figure \ref{fig:error_simple_quadrupole}, concerning the full computational domain $\Omega$ and the individual patches $\Omega_1, \dots, \Omega_4$.
Even though a low absolute error is obtained throughout the domain, the maximal error occurs in the subdomain $\Omega_1$, in particular in areas where the \gls{nurbs} parametrization exhibits non-smooth behavior.
This is also observed in Figure~\ref{fig:contour_simple_quadrupole}, which shows the absolute error over the computational domain.
The minimum relative $L^2$ error over $\Omega$ is $1.77\cdot 10^{-2}$.

\subsubsection{Complex quadrupole magnet model}
\label{subsec:num_prob2}

\begin{figure}[t!]
\centering
\includegraphics[width = 0.45\linewidth]{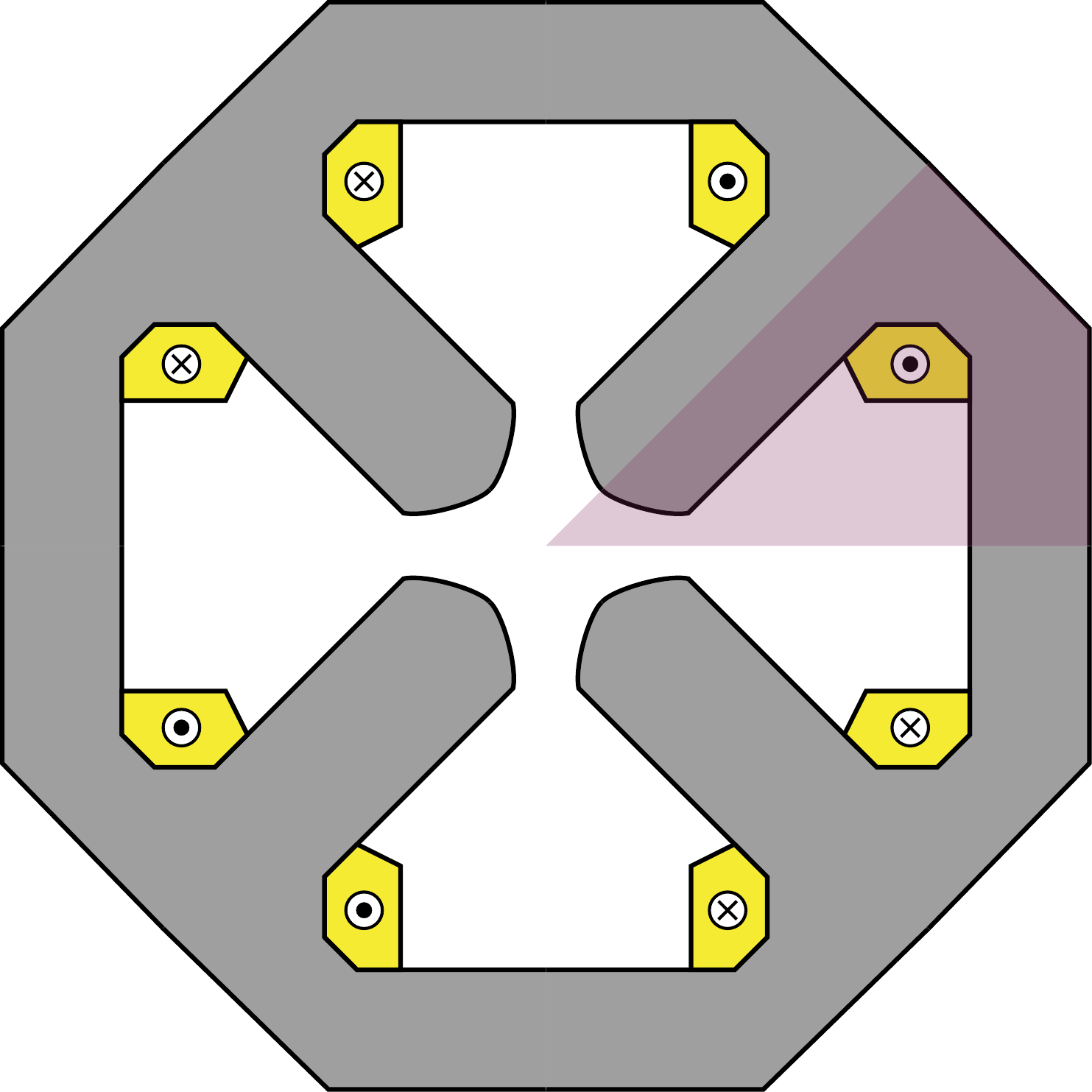}
\includegraphics[width = 0.5\linewidth]{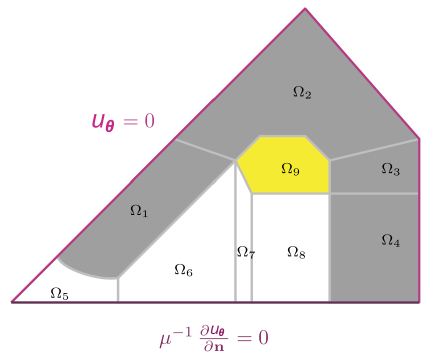}
\caption[Complex quadrupole model geometry]{Domain geometry of the complex quadrupole model. The iron yoke is shown in gray, the current excitations in yellow, and the air gap in white. The purple shading indicates the model section in the symmetry planes. \textbf{Left:} Full cross-section. 
\textbf{Right:} One-eighth of cross-section, exploiting rotational symmetry. The domain is partitioned into patches $\Omega_i$, $i=1, \dots, 9$.}
\label{fig:prob1_complex_quad}
\end{figure}

Next, we consider a more complex quadrupole magnet geometry,  depicted in Figure \ref{fig:prob1_complex_quad}.
The left hand side of Figure \ref{fig:prob1_complex_quad} shows the complete cross-section of the quadrupole magnet, while the right hand side shows one eighth of its geometry, exploiting its rotational symmetry. 
The latter constitutes the domain geometry of the computational model and is partitioned into patches $\Omega_1,...,\Omega_9$.
The material distribution and coloring scheme is identical to Figure \ref{fig:geo_simple_quad}.
There are 5 more \gls{nurbs} parametrizations than in the previous section, hence the characterization ``complex''.
The ansatz function for this example is more intricate, as conformity must be ensured over nine patches instead of four.
Again, Dirichlet and Neumann boundary conditions are applied at the boundaries as indicated.

Again, a \gls{resnet} architecture is employed for both the domain- and interface-\glspl{nn}.
In this case, we employ \glspl{nn} with five \gls{fc} layers, including an input layer, one \gls{resnet} block, and an output layer.
For the domain-\glspl{nn}, we employ two variations of the same architecture.
In subdomains $\Omega_1$ and $\Omega_2$, we allocate $16$ neurons per layer, such that the input and output dimensions for each \gls{fc} layer vary throughout the network according to $(2,16) \rightarrow (16,16) \rightarrow (16,16) \rightarrow (16,16) \rightarrow(16,1)$.
For the rest subdomains, we allocate eight neurons per layer, such that the input and output dimensions vary according to $(2,8) \rightarrow(8,8) \rightarrow(8,8) \rightarrow(8,8) \rightarrow(8,1)$.
For the interface-\glspl{nn}, we employ five neurons per layer and the input and output dimensions for each \gls{fc} layer vary according to $(1,5) \rightarrow(5,5) \rightarrow(5,5) \rightarrow(5,5) \rightarrow(5,1)$.
In both cases there are two skip connections, before the third and the final layer, respectively.
The activation function used is the hyperbolic tangent function $\sigma(x) = \tanh(x)$. 
The \glspl{nn} are trained for 1000 epochs, utilizing $ 1.8 \times 10^5 $ Monte Carlo sampling points per epoch. 
For the reference solution, we employ the FEniCS \gls{fem} solver \cite{AlnaesEtal2014, A.Logg.2010}, using a first-order tetrahedral mesh consisting of 770743 elements. %
Table \ref{table:train_quad_complex} we provides an overview of the \gls{nn} and trainable parameters employed in this use-case.

\begin{table}[b!]
	\caption{Distribution of \glspl{nn} and trainable parameters for the complex quadrupole model.}
	\centering
	\begin{tabular}{l c c}
		\hline\hline
		  & Number of \glspl{nn} &Number of trainable parameters \\ [0.5ex] 
		\hline
		Point interfaces (0D) 						& --&5 		 	\\[0.5ex] 
		Interface \glspl{nn} (1D) 			  			&13&74 			\\[0.5ex] 
		  Domain \glspl{nn} for $\Omega_3$ -- $\Omega_9$ (2D) 	 		&7&183		 	\\[0.5ex] 
		Domain   \glspl{nn} for $\Omega_1$, $\Omega_2$ (2D)					&2&615 	 	\\[0.5ex] 
		Total									& 22 & 3478	 	\\[0.5ex] 
		\hline
	\end{tabular}
	\label{table:train_quad_complex}
\end{table}

\begin{figure}[t!]
\centering
\includegraphics[width = 0.8\linewidth]{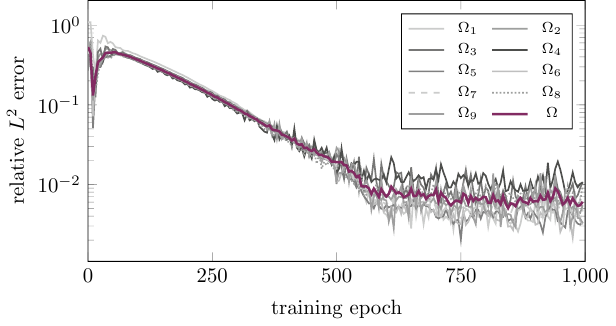}
\caption{Relative $L^2$ error during training for the entire computational domain and for each subdomain and the complex quadrupole model.}
\label{fig:error_complex_quadrupole}
\end{figure}

\begin{figure}[t!]
\centering
\begin{subfigure}{0.49\textwidth}
    \centering
    \includegraphics[width=\linewidth]{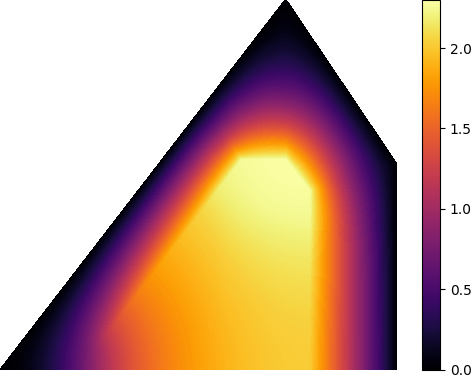}
    \subcaption{\gls{pinn} solution.}
    \label{fig:first}
\end{subfigure}
\hfill
\begin{subfigure}{0.49\textwidth}
    \centering
    \includegraphics[width=\linewidth]{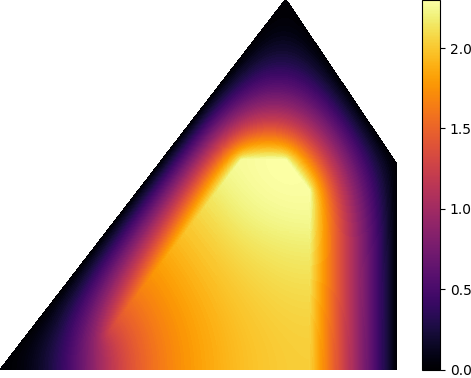}
        \subcaption{Reference \gls{fem} solution.}
        \label{fig:second}
\end{subfigure}
\begin{subfigure}{0.49\textwidth}
    \centering
    \includegraphics[width=\linewidth]{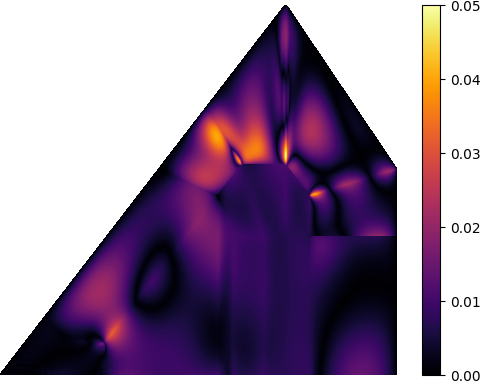}
    \subcaption{Pointwise error.}
    \label{fig:third}
\end{subfigure} 
\caption{Field solutions and corresponding error field for the complex quadrupole magnet model.}
\label{fig:error_contour}
\end{figure}

Similar to the previous section, the \glspl{nn} are trained for 1000 training epochs and the relative $L^2$ error is calculated with respect to an \gls{fem} reference solution using $4\cdot10^4$ quadrature points.
The relative $L^2$ error over the entire training process is displayed in Figure \ref{fig:error_complex_quadrupole} for both the full computational domain and its subdomains.
Although the absolute error is consistently low across the domain, the highest error occurs within the subdomain $\Omega_2$, particularly in areas where the \gls{nurbs} parameterization displays non-smooth behavior.
This is also evident in Figure \ref{fig:third}, which shows of the absolute pointwise error over the quadrupole's computational domain.
The minimum relative $L^2$ error over $\Omega$ is $6.05\cdot 10^{-3}$. 

\subsection{Solid mechanics simulation of mechanical holder}
\label{sec:num_example_mechanics}

\subsubsection{Solid mechanics formulation}
\label{sec:solid_mechanics_formulation}

We now consider a problem in the setting of nonlinear solid mechanics. 
Given a domain $\Omega \subset \mathbb{R}^d$, $d=2,3$, composed of $N_\Omega$ patches, we define the displacement field $\bm{u} : \Omega \rightarrow \mathbb{R}^d$, which describes the deformation of the domain $\Omega$. 
Then, given a point $\mathbf{x} \in \Omega$, its position in the deformed configuration is given by $\bm{\varphi}(\mathbf{x}) = \mathbf{x} + \bm{u}(\mathbf{x})$.
The task is to find the displacement that minimizes the total potential energy \cite{bonet2021nonlinear}:
\begin{align}
    \min\limits_{\bm{u} \in \mathcal{V}} I(\bm{u}) = \min\limits_{\bm{u}} \sum\limits_{i=1}^{N_\Omega} \left(\int\limits_{\Omega_i} \psi(\bm{u}) \text{ d} \mathbf{x}  - \int\limits_{\Omega_i} \bm{b} \cdot \bm{u} \text{ d} \mathbf{x} -  \int\limits_{\Gamma_i} \bm{T} \cdot \bm{u} \text{ d} \sigma(\mathbf{x}) \right), \label{eq:energy_solid_mechanics}
\end{align}
where $\mathcal{V}$ is a suitable space of all possible displacements satisfying the boundary conditions, $\psi$ is the stored energy density, $\bm{b}$ is the body force, and $\bm{T}$ is the surface traction prescribed on some parts of the boundary $\Gamma_i \subset \partial\Omega, i=1,...,N_\Omega$. 
The boundaries are considered to be the union of images of facets in the reference domain, such that $\Gamma_i=\bm{f}_{i}(\hat{\Gamma}_i)$. Both body forces and surface traction are defined in the undeformed configuration. In this work we use the Saint Venant–Kirchhoff material model \cite{bonet2021nonlinear} with the stored energy density
\begin{align}
    \psi = \frac{\lambda}{2} (\text{tr} \bm{E})^2 + \mu \bm{E} \cdot \bm{E}, \label{eq:sv_model}
\end{align}
where $\bm{E} = \frac{1}{2}(\bm{F} \bm{F}^\top - \bm{I})$ is the Green strain tensor, $\bm{F} = \bm{I} + \textbf{grad} \bm{u}$ is the deformation gradient and $\lambda$, $\mu$ are the Lam\'e coefficients. The formulation can be expressed in the reference domains $\hat{\Omega}_i$ as
\begin{align}
    I(\hat{\bm{u}}) =  \sum\limits_{i=1}^{N_\Omega} \left(\int\limits_{\hat{\Omega}_i} \hat{\psi}(\hat{\bm{u}}) | \text{det} D \bm{f}_{i} | \text{ d} \hat{\mathbf{x}}  - \int\limits_{\Omega_i} \bm{b} \cdot \hat{\bm{u}} \:  | \text{det} D \bm{f}_{i} | \text{ d} \hat{\mathbf{x}} -  \int\limits_{\Gamma_i} \bm{T} \cdot \bm{u} ||D \bm{f}_{i} \bm{\nu} || |\text{det} D \bm{f}_{i}| \text{ d} \sigma(\hat{\mathbf{x}}) \right), \label{eq:energy_solid_mechanics_iga}
\end{align}
where $\hat{\psi} = \frac{\lambda}{2} (\text{tr} \hat{\bm{E}})^2 + \mu \hat{\bm{E}} \cdot \hat{\bm{E}}$, 
$\hat{\bm{E}} = \frac{1}{2}(\hat{\bm{F}} \hat{\bm{F}}^\top - \bm{I})$ and $\hat{\bm{F}} = \bm{I}+(\textbf{grad } \hat{\bm{u}} ) (D \bm{f}_{i})^{-1}$ and $\bm{\nu}_i, i=1,...,N_\Omega$ are the unit normals.

\begin{figure}
    \centering
    \begin{tikzpicture}[scale=1.2]

\draw[thick] (-2,0) -- (-0.5,0) -- (-0.5,-0.5) -- (-2,-0.5) -- cycle;
\node at (-1.25,-0.7) {$\Omega_r$};

\draw[thick] (-1.25,1.2) circle (0.6);
\node at (-1.25,1.2) {$\bm{\varphi}(\Omega)$};

\draw[dotted] (-1.25,0.6) -- (-1.25,0);
\node at (-0.7,0.3) {$g_{\Omega^{\rm r}}=0$};

\draw[thick] (1,0) -- (2.5,0) -- (2.5,-0.5) -- (1,-0.5) -- cycle;
\node at (1.75,-0.7) {$\Omega_r$};

\draw[thick] (1.75,0.4) circle (0.6);
\node at (1.75,0.4) {$\bm{\varphi}(\Omega)$};

\draw[dotted] (1.75,0.0) -- (1.75,-0.2);
\node at (3.2,0.0) {$g_{\Omega^{\rm r}}>0$};

\end{tikzpicture}
    \caption{Illustration of the penetration distance between a rigid body $\Omega^{\rm r}$ and the deformed configuration $\bm{\varphi}(\Omega)$. On the left, the penetration distance is zero as no penetration yet exists. On the right, the penetration distance becomes strictly positive.}
    \label{fig:penetration}
\end{figure}
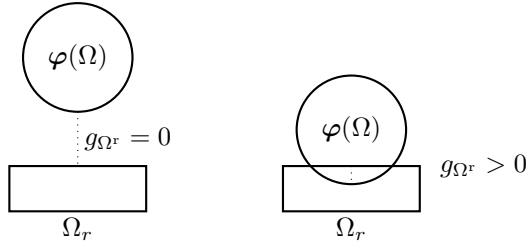

A further topic of interest in solid mechanics simulations are contact boundary conditions. 
In this work, we consider a second body, denoted as $\Omega^{\rm r} \subset \mathbb{R}^d$, which is  rigid and not part of the computational domain (see Figure \ref{fig:penetration}).
To prevent the penetration of the two bodies, we use the penalty approach \cite{wriggers2006computational}. 
The idea is to apply a traction along the boundary subdomain where the penetration occurs. 
To that end, we define the penetration distance function from a point $\bm{\varphi}(\mathbf{x}), \mathbf{x} \in \partial \Omega$ to the body $\Omega^{\rm r}$ as
\begin{align}
    g_{\Omega^{\rm r}}(\bm{\varphi}, \mathbf{x}) = \begin{cases}
        \min\limits_{\bm{x'} \in \Omega^{\rm r}} ||\bm{\varphi}(\mathbf{x}) - \mathbf{x}'||_2^2, & \quad \bm{\varphi}({\mathbf{x}}) \text{ penetrates } \Omega^{\rm r},\\
                0, & \quad\text{otherwise}.
    \end{cases}
\end{align}

The following penalty is introduced:
\begin{align}
I_{\rm C} =
\frac{1}{2} \int\limits_{\partial\Omega}\epsilon_{\rm N} g_{\Omega^{\rm r}}^2(\bm{\varphi}, \mathbf{x})  \: ||\bm{F}^{-\top} \bm{N} ||_2  \: \det \bm{F} \: \text{ d}\sigma (\mathbf{x}),  \label{eq:pentaly_integrals} \end{align}
where $\bm{N}$ is the normal vector on the undeformed boundary and $\epsilon_{\rm N} > 0$ is a penalty constant. The integral can be expressed in the reference configuration using \eqref{eq:energy_solid_mechanics_iga}. By adding the penalty and the potential energy and introducing the conformal ansatz functions, denoted by $\bm{u}_{\bm{\theta}}$, the final optimization problem to be solved is: 
\begin{align}
    \min\limits_{\bm{u}_{\bm{\theta}} \in \mathcal{V}} I(\bm{u}_{\bm{\theta}}) + I_{\rm C}(\bm{u}_{\bm{\theta}}).
\end{align}

The penalty term expressed in \eqref{eq:pentaly_integrals} is also nonlinear with respect to the displacement. Conventional FEM based methods employ Newton's method to cope with the nonlinearity. 
In the case of nonlinear solid mechanics with contact boundary conditions, incremental loading (displacement applied on boundary, body forces or boundary traction) is used to avoid the divergence of the Newton iterations due to badly selected starting points, and the previous increment is used as a starting point for the next. 
In the framework developed in this work, we solve the stationary problem directly. 
The load can be added as an additional input to the \glspl{nn} that approximate the solution. 

\subsubsection{Mechanical holder model}

\begin{figure}[b!]
	\centering
	\includegraphics[width=\textwidth]{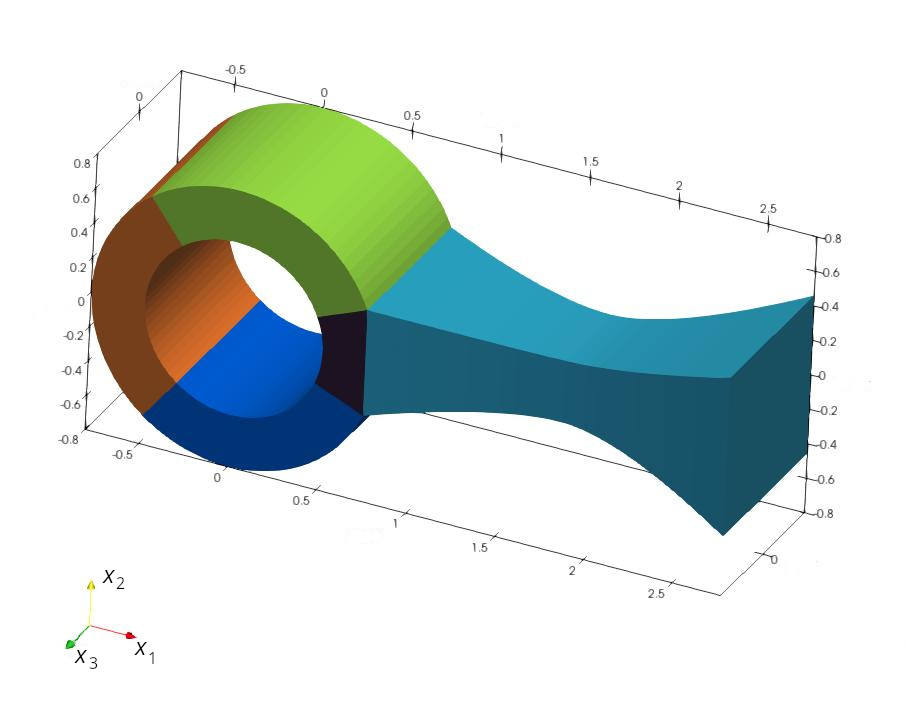}
	\caption{Geometry of the mechanical holder model, parametrized with five NURBS patches, each shown with a different color. All dimensions are expressed in mm.}
	\label{fig:holder_geometry}
\end{figure}

In this section, we examine a structural holder, the geometry of which is depicted in Figure \ref{fig:holder_geometry}. 
The computational domain consists of five NURBS patches, also shown in Figure \ref{fig:holder_geometry}. 
To model the material behavior, we utilize the Saint Venant-Kirchhoff constitutive law presented in \eqref{eq:sv_model}, with the Lam\'e parameters $\lambda=\frac{E \nu}{(1+\nu)(1-2 \nu)}$ and $\mu = \frac{E}{2(1+\nu)}$, where the Young's modulus is $E=2000$ MPa and the Poisson's ratio is $\nu=0.3$. 

The mechanical setup includes two rigid half-planes, defined by the relations $x_2 = -0.75 - 0.05\phi_1$ for $x_1 \leq 1$, and $x_2 = 0.75 + 0.05\phi_2$ for $x_1 \leq 1$, which make contact with the domain. The variables $\phi_1, \phi_2 \in [-1,1]$ serve as additional inputs to the \glspl{nn} that control the boundary conditions. 
The object is fixed at $x_1=2.75$ by imposing a zero Dirichlet boundary condition on the displacement.

The solution is approximated using a total of twelve \glspl{nn}, where five \glspl{nn} are employed for the five subdomains, five \glspl{nn} are defined on the surface interfaces between the domains, and two \glspl{nn} are defined on the line interfaces where three domains meet. All \glspl{nn} have the same architecture, consisting of a single fully connected input layer, followed by two residual blocks. 
    Each residual block contains two hidden layers with 16 neurons each, thus resulting in a total of $14676$ trainable parameters. The layer by layer structure is $(d+2,16)\rightarrow[(16,16)\rightarrow(16,16)]\rightarrow[(16,16)\rightarrow(16,16)] \rightarrow (16, 3)$, where $d$ is the dimensionality of the manifold the network is defined on and $2$ stands for the number of parameters.
A detailed overview is given in Table \ref{tab:holder_params}.
The activation function used is the squared ReLU $\sigma(x) = \max(x^2, 0)$. 
The \glspl{nn} are trained for 64 epochs, utilizing $ 8 \times 10^6 $ Monte Carlo sampling points per epoch, divided into $1000$ batches. 
The training process is executed on an Nvidia RTX 4090 laptop GPU, with a total runtime of 13 minutes. 
For the reference solution, we employ the CalculiX finite element solver \cite{dhondt2004finite}, using a first-order tetrahedral mesh consisting of 770743 elements. 

\begin{table}[b!]
	\caption{Distribution of \glspl{nn} and trainable parameters for the mechanical holder model.}
	\centering
	\begin{tabular}{l c c}
		\hline\hline
		   & Number of \glspl{nn} &Number of trainable parameters \\ [0.5ex] 
		\hline
		Domain \glspl{nn} (3D)						& 5 &  1235 		 	\\[0.5ex] 
		Interface \glspl{nn} (2D) 			  			& 5 &1219 			\\[0.5ex] 
            Interface \glspl{nn} (1D) 			  			& 2 &1203 			\\[0.5ex] 
		Total 									& 12 & 14676	 	\\[0.5ex] 
		\hline
	\end{tabular}
	\label{tab:holder_params}
\end{table}

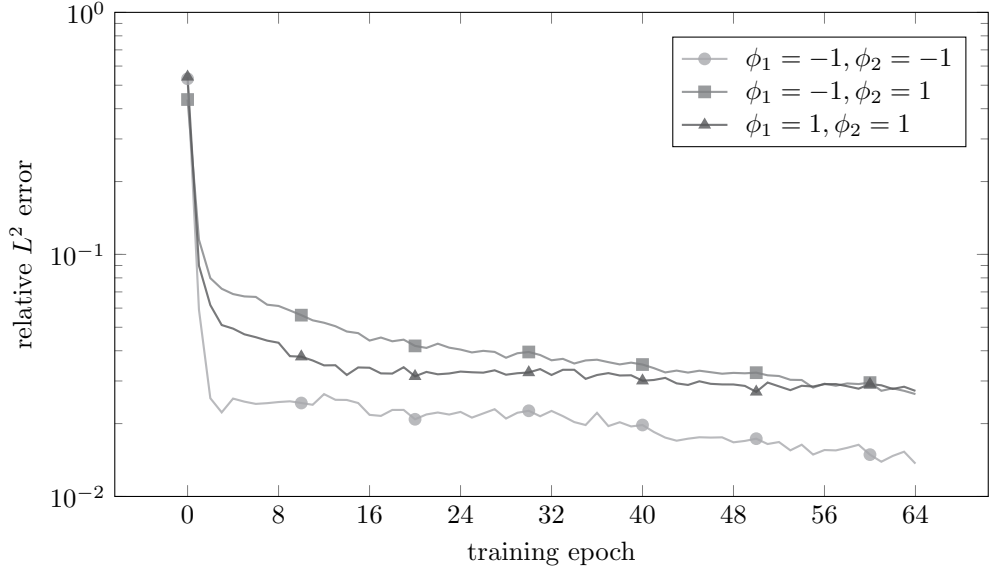
\begin{figure}
    \centering
\begin{tikzpicture}
\definecolor{TUDa-0d}{cmyk/RGB/HTML}{0,0,0,.8/83,83,83/535353}
\definecolor{TUDa-0c}{cmyk/RGB/HTML}{0,0,0,.6/137,137,137/898989}
\definecolor{TUDa-0b}{cmyk/RGB/HTML}{0,0,0,.4/181,181,181/B5B5B5}
\definecolor{TUDa-0a}{cmyk/RGB/HTML}{0,0,0,.2/220,220,220/DCDCDC}
\definecolor{TUDa-10c}{cmyk/RGB/HTML}{.5,1,.3,0/149,17,105/951169}
\definecolor{darkgray153}{RGB}{153,153,153}
\definecolor{darkgray176}{RGB}{176,176,176}
\definecolor{darkslategray45}{RGB}{45,45,45}
\definecolor{dimgray102}{RGB}{102,102,102}
\definecolor{lightgray204}{RGB}{204,204,204}

\begin{semilogyaxis}
[
		grid=none,
		width = \textwidth,
		height = 8cm,
            ymin=0.01,
            ymax=1,
            scaled ticks = false,
		xtick = {0,8,16,24,32,40,48,56,64},
            ytick={0.01,0.1,1},
            x tick label style={/pgf/number format/fixed,/pgf/number format/precision=0},
            y tick label style={ /pgf/number format/fixed, /pgf/number format/precision=3},
		xlabel={training epoch},
		ylabel={relative $L^2$ error},
		legend cell align={left},
            legend style={at ={(0.975,0.95)}, fill=none, column sep = 1ex},
		]

]
\addplot [thick, TUDa-0b, opacity=0.8, mark=*, mark size=2, mark repeat=10, mark options={solid}]
table {%
0 0.531621694161907
1 0.05911375112171
2 0.0254644547557862
3 0.0222503148050494
4 0.0253722741424082
5 0.0246408574780716
6 0.0241223767476214
7 0.024291281369668
8 0.024554484226631
9 0.0247229098659408
10 0.0243271055904179
11 0.023921432344874
12 0.0264752734175456
13 0.025080188138933
14 0.0250428671103537
15 0.024322794585933
16 0.0217357639299421
17 0.0215186916877016
18 0.0227725457313128
19 0.0227787557367247
20 0.0208377316523218
21 0.0218186772858094
22 0.0222244115866901
23 0.0218013183668544
24 0.0223515986335583
25 0.0211719736825258
26 0.0220341302171116
27 0.0229324421547224
28 0.0209884871150468
29 0.0221730186820669
30 0.0225942623300265
31 0.0214497542604759
32 0.0225634487227285
33 0.0214753429352307
34 0.0202913371279953
35 0.0197304666626872
36 0.022153795897068
37 0.019532793556463
38 0.0202588921608499
39 0.0194690872935086
40 0.0197360224135301
41 0.0184677976844481
42 0.0175064785979027
43 0.0169976595617041
44 0.0173261872622821
45 0.0175623929331732
46 0.0175096654612465
47 0.0175494341682918
48 0.0167385570006671
49 0.0169609076488985
50 0.0173269979882898
51 0.0164698419259005
52 0.0167848375357953
53 0.0155075309312515
54 0.0163584685126864
55 0.01489719149566
56 0.015553518196882
57 0.0155006831051189
58 0.0158826303410272
59 0.01635170320162
60 0.0148857358606049
61 0.0139152979838548
62 0.0147036404970005
63 0.0153087508751875
64 0.0136700000000001
};
\addlegendentry{$\phi_1=-1, \phi_2=-1$}
\addplot [thick, TUDa-0c, opacity=0.8, mark=square*, mark size=2, mark repeat=10, mark options={solid}]
table {%
0 0.436427331882468
1 0.11495039640413
2 0.0798160019746398
3 0.0721464222662477
4 0.0685799753181199
5 0.0670552394594549
6 0.0667354624517355
7 0.0619809614431685
8 0.0612406399895461
9 0.058678225882314
10 0.0561126601900682
11 0.0534185888803632
12 0.0521988711037068
13 0.0505077104494435
14 0.0480256514753199
15 0.0472930683462213
16 0.0440587694510914
17 0.0453494142134571
18 0.0437651267265459
19 0.0444457040876098
20 0.0418841538387978
21 0.0410886672227706
22 0.0427734894518929
23 0.0411927219087745
24 0.0404477653943172
25 0.039367110160751
26 0.0399582526734223
27 0.0395572845922737
28 0.0374226253237815
29 0.039059787526892
30 0.0395258824182666
31 0.0384124766734621
32 0.0365299471167626
33 0.0370274834881513
34 0.0354257244937631
35 0.0364741790255476
36 0.0367282923922775
37 0.0358860013010525
38 0.0351244976592614
39 0.0357220967588122
40 0.0350784877391667
41 0.0338986538592655
42 0.0325445572378475
43 0.0331105096810985
44 0.0324993473316688
45 0.0330793310666897
46 0.0326106257229357
47 0.032145245102653
48 0.0323995427992092
49 0.0322726477270129
50 0.0324665152241917
51 0.0316656268060387
52 0.0314537099543714
53 0.0303397571216842
54 0.030276089594571
55 0.028159151099087
56 0.0292251757400574
57 0.0286725890058001
58 0.0293099714371193
59 0.0291341915115165
60 0.0295442991036489
61 0.0273260838729965
62 0.0279710385631672
63 0.0273145179980913
64 0.0264800000000001
};
\addlegendentry{$\phi_1=-1, \phi_2=1$}
\addplot [thick, TUDa-0d, opacity=0.8, mark=triangle*, mark size=2, mark repeat=10, mark options={solid}]
table {%
0 0.541411325690769
1 0.089788753076458
2 0.0617593887340795
3 0.0510740492989348
4 0.049369361739333
5 0.0467549046828026
6 0.0454931431132556
7 0.044042001030383
8 0.0431563250796637
9 0.0379915454807225
10 0.0378059348719205
11 0.036460000278905
12 0.0348028768374863
13 0.0348430873065727
14 0.0317658836332426
15 0.0341167105419881
16 0.0340250538391538
17 0.0322465663956023
18 0.0321880015542197
19 0.0341887052151518
20 0.0313686017919562
21 0.0326999314079183
22 0.0319720631787421
23 0.0322198839943553
24 0.0328056256416563
25 0.0325933863713564
26 0.0324723380726569
27 0.0332047658296489
28 0.0318439200069037
29 0.0322944578646004
30 0.0325298144390535
31 0.033596233734102
32 0.0317598570928278
33 0.0333690191143387
34 0.0333713156961108
35 0.0306075734060827
36 0.0317519811535283
37 0.0323286898070443
38 0.0316058720818511
39 0.0316702342904759
40 0.0300779906692602
41 0.0302890622177414
42 0.0308966925492433
43 0.0293270894209876
44 0.02888822738181
45 0.0299108807896903
46 0.0291968459166909
47 0.0290337792109815
48 0.0290166660719665
49 0.0286273420204488
50 0.0270504599194226
51 0.0295569226521446
52 0.0285014533887604
53 0.0274962758375115
54 0.028659077716779
55 0.0284646139916486
56 0.0291067428582599
57 0.0291362395438774
58 0.0285421319707479
59 0.0278770470027934
60 0.0291224999297078
61 0.0287958927565543
62 0.0278146327087704
63 0.0284013900760487
64 0.0273000000000001
};
\addlegendentry{$\phi_1=1, \phi_2=1$}
\end{semilogyaxis}

\end{tikzpicture}
    \caption{Relative $L^2$ error during training for the entire computational domain and for different combinations of the parameters $\phi_1, \phi_2$.}
    \label{fig:solid-mechanics-epochs}
\end{figure}

\begin{figure}[t!]
\centering
\begin{subfigure}{0.49\textwidth}
    \centering
    \includegraphics[width=\textwidth]{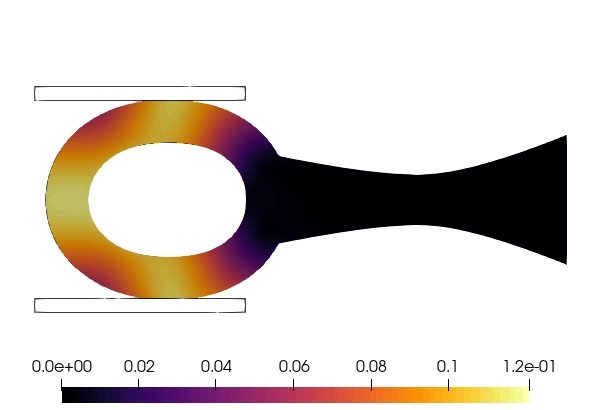}
    \caption{PINN solution for $\phi_1=-1, \phi_2=-1$.}
\end{subfigure}
\begin{subfigure}{0.49\textwidth}
    \centering
    \includegraphics[width=\textwidth]{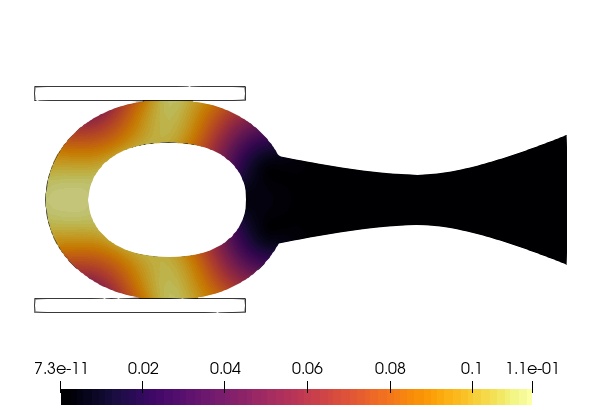}
    \caption{Reference FEM solution for $\phi_1=-1, \phi_2=-1$.}
\end{subfigure}
\begin{subfigure}{0.49\textwidth}
    \centering
    \includegraphics[width=\textwidth]{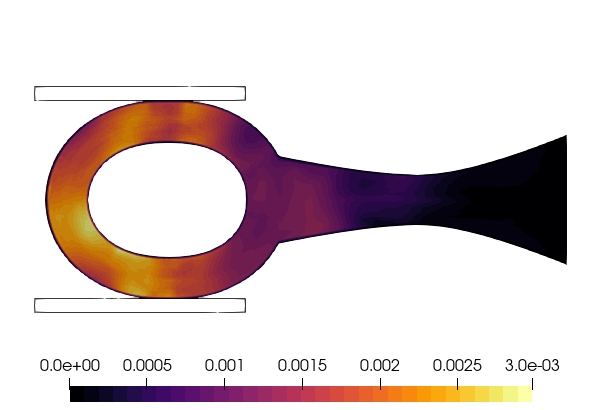}
    \caption{Pointwise error for $\phi_1=-1, \phi_2=-1$.}
\end{subfigure} 
\caption{Field solutions and corresponding error field for the mechanical holder model with contact parameters $\phi_1=-1$, $\phi_2=-1$. 
A 2D slice at $ x_3 = 0 $ is used for the visualization.}
\label{fig:deformed_comparison_0}
\end{figure}

\begin{figure}[t!]
\centering
\begin{subfigure}{0.49\textwidth}
    \centering
    \includegraphics[width=\textwidth]{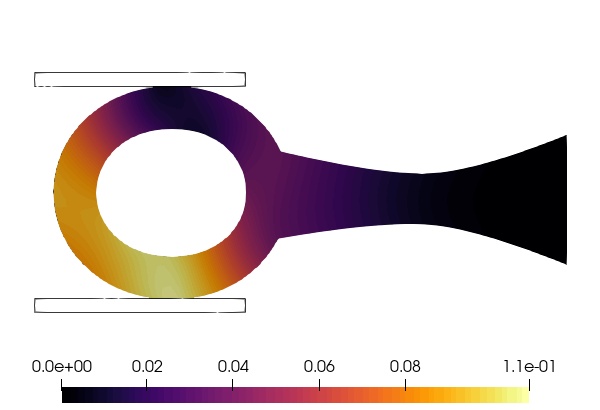}
    \caption{PINN solution for $\phi_1=-1, \phi_2=1$.}
\end{subfigure}
\begin{subfigure}{0.49\textwidth}
    \centering
    \includegraphics[width=\textwidth]{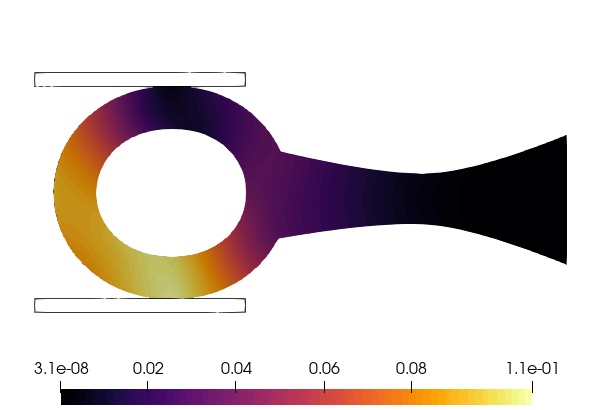}
    \caption{Reference FEM solution for $\phi_1=-1, \phi_2=1$.}
\end{subfigure}
\begin{subfigure}{0.49\textwidth}
    \centering
    \includegraphics[width=\textwidth]{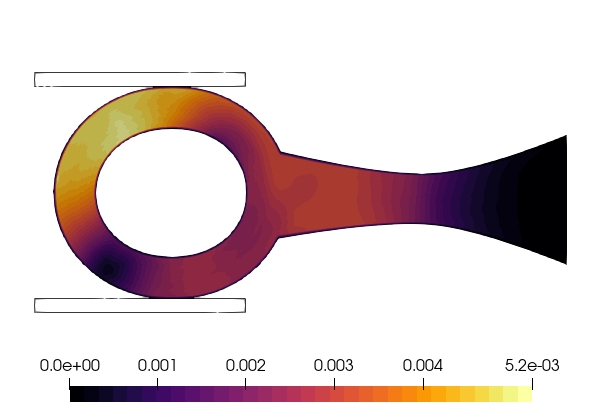}
    \caption{Pointwise error for $\phi_1=-1, \phi_2=1$.}
\end{subfigure} 
\caption{Field solutions and corresponding error field for the mechanical holder model with contact parameters $\phi_1=-1$, $\phi_2=1$. 
A 2D slice at $ x_3 = 0 $ is used for the visualization.}
\label{fig:deformed_comparison_1}
\end{figure}

\begin{figure}[t!]
\centering
\begin{subfigure}{0.49\textwidth}
    \centering
    \includegraphics[width=\textwidth]{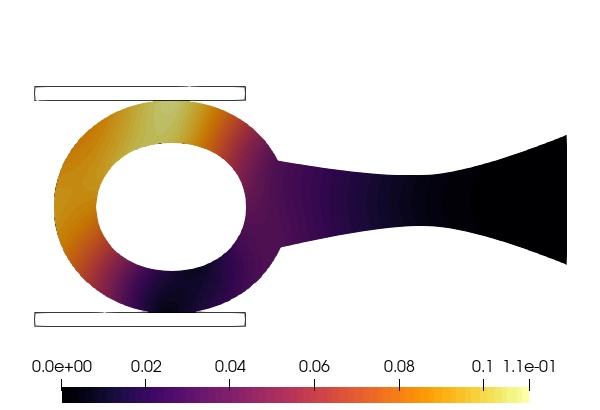}
    \caption{PINN solution for $\phi_1=1, \phi_2=-1$.}
\end{subfigure}
\hfill
\begin{subfigure}{0.49\textwidth}
    \centering
    \includegraphics[width=\textwidth]{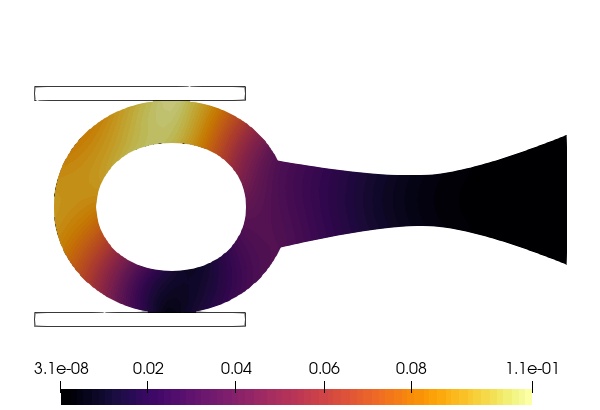}
    \caption{Reference FEM solution for $\phi_1=1, \phi_2=-1$.}
\end{subfigure}
\begin{subfigure}{0.49\textwidth}
    \centering
    \includegraphics[width=\textwidth]{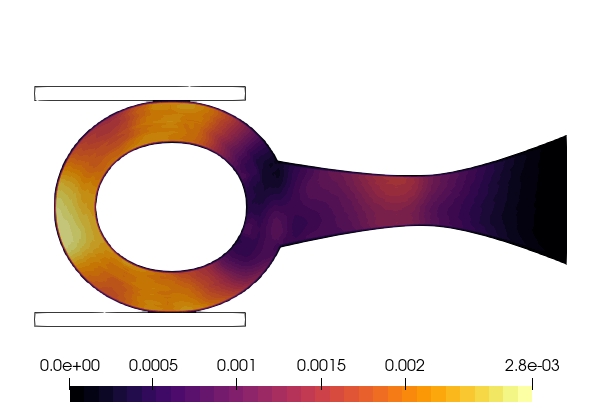}
    \caption{Pointwise error for $\phi_1=1, \phi_2=-1$.}
\end{subfigure} 
\caption{Field solutions and corresponding error field for the mechanical holder model with contact parameters $\phi_1=1$, $\phi_2=-1$. 
A 2D slice at $ x_3 = 0 $ is used for the visualization.}
\label{fig:deformed_comparison_2}
\end{figure}

The results are presented in Figures \ref{fig:deformed_comparison_0}-\ref{fig:deformed_comparison_2}, where both the deformed configurations and the pointwise error fields are shown for the parameter combinations $(\phi_1, \phi_2) \in \{(-1,-1), (-1,1), (1,-1)\}$. 
All shown results correspond to a 2D slice at $x_3 = 0$. 
The case $(\phi_1, \phi_2) = (1,1)$ represents a configuration with no contact and where no deformation exists.
As a parameter dependent error metric, we use the error metric defined in \eqref{eq:rel-L2-error},
where $\bm{u}$ represents the predicted displacement field and $\bm{u}^{\rm FEM}$ is the reference solution. 
The error is evaluated over a Cartesian grid defined on each individual NURBS patch. 
The number of grid points is $N=75000$, corresponding to a uniform grid on the reference domain.
The error values for the different parameter combinations of $\phi_1$ and $\phi_2$ are given in Table \ref{tab:error_values}. 
The behavior of the error during training is shown in Figure \ref{fig:solid-mechanics-epochs}.

\begin{table}[b!]
    \centering
    \begin{tabular}{ccc}
        \hline
        $\phi_1$ & $\phi_2$ & $ \epsilon_{\text{rel}, L^2}$ \\
        \hline
        -1 & -1 & 0.0136 \\ 
        -1 & 1  & 0.0264 \\ 
        1  & -1 & 0.0266 \\ 
        \hline
    \end{tabular}
    \caption{Relative $L^2$ error for different combinations of $\phi_1$ and $\phi_2$.}
    \label{tab:error_values}
\end{table}

\section{Conclusion}
\label{sec:conclusion}

This paper presented a novel methodology combining \glspl{pinn} with the \gls{iga} framework tailored for complex, multi-patch domains common in \gls{cad}. 
Our core contribution lies in formulating the \gls{pinn} approximation within the reference domain of individual \gls{iga} patches, ensuring geometric fidelity. 
Solution continuity across patch interfaces was achieved through the introduction of dedicated interface \glspl{nn}, while the strong enforcement of Dirichlet boundary conditions was realized via a specific \gls{nn} output ansatz, drawing parallels to domain decomposition techniques similar to \gls{feti} and mortaring methods.

The efficacy and versatility of the developed multi-patch isogeometric neural solver was validated through complex numerical examples related to real-world applications. 
We successfully applied the framework to both 2D magnetostatic simulations, involving simple and complex quadrupole magnet geometries, and a 3D nonlinear solid mechanics problem featuring hyperelastic material behavior, contact boundary conditions, and parametric dependencies. Quantitative numerical error analyses against reference solutions demonstrated the accuracy and robustness of the proposed method.

This work contributes towards bridging the gap between native \gls{cad} representations and mesh-free simulation capabilities offered by \glspl{pinn}. 
By effectively handling complex multi-patch geometries and ensuring strong enforcement of essential boundary and interface conditions, our framework overcomes key limitations of standard \glspl{pinn} and provides a powerful tool for analyzing complex physical systems directly within their design context.

Naturally, the suggested neural solver comes with a set of limitations, which we plan to address in future studies. 
First, suitable \gls{nn} architectures are selected empirically and by trial and error.
Since the focus of this work is on embedding \glspl{pinn} into the \gls{iga} framework, a systematic study on specific \gls{nn} architectures remains out of scope, leaving many questions regarding design and optimization open for future research. 
The neural solver is also not yet competitive to traditional numerical solution methods in terms of computational efficiency, due to the required \gls{nn} training times. 
Limitations related to those encountered in conventional \gls{iga} also exist.
For example, the suggested neural solver requires interface-conforming patches, for which no automatic generation procedure exists, even in conventional \gls{iga}. 
Moreover, the imposition of the Dirichlet boundary condition still lacks the Kronecker delta property, which is a known limitation of \gls{iga} compared to classical \gls{fem}.
Last, the neural solver is currently limited to enforcing $C^0$-continuity only. 
Extensions to higher regularities remain to be addressed.

Last, in addition to resolving the aforementioned limitations, the generalization of the suggested neural solver to parametric problem formulations, in particular high-dimensional ones, constitutes an open and highly relevant future research direction.
From the perspective of the authors, research on parametric problems that leverages optimized, problem-specific \gls{nn} architectures holds significant potential to surpass conventional methods, similar to the successes of \glspl{nn} in fields such as image classification and natural language processing.

\bmhead{Author contributions}
MvT and IGI developed the methodology, implemented the software, and performed all simulations and numerical studies. DL acquired funding and provided supervision. All authors contributed to writing and reviewing the manuscript. 


\bibliography{sn-bibliography}

\end{document}